\documentclass[hidelinks,letterpaper,twocolumn,10pt]{article}
\usepackage{usenix-2020-09}

\usepackage{tikz}
\usepackage{amsmath}
\usepackage{amssymb}
\usepackage{pifont}
\usepackage{glossaries}
\usepackage{paralist}
\usepackage{xcolor,colortbl}
\usepackage{tabularx,ragged2e,booktabs,caption}

\usepackage{enumitem}
\usepackage{nameref}

\newlist{questions}{enumerate}{2}
\setlist[questions,1]{label=RQ\arabic*.,ref=RQ\arabic*}
\setlist[questions,2]{label=(\alph*),ref=\thequestionsi(\alph*)}

\usepackage{mathtools,eqparbox}

\newcommand{\xmark}{\text{\ding{55}}}
\newcolumntype{P}[1]{>{\centering\arraybackslash}p{#1}}

\makeglossaries

\newglossaryentry{pm}
{
    name=PM,
    description={Persistent Memory}
}

\newglossaryentry{dram}
{
    name=DRAM,
    description={Dynamic Random-Access Memory}
}

\newglossaryentry{fs}
{
    name=FS,
    description={File System}
}

\newglossaryentry{kv}
{
    name=KV,
    description={Key-Value}
}

\newglossaryentry{dax}
{
    name=DAX,
    description={Direct Access}
}

\newglossaryentry{ssd}
{
    name=SSD,
    description={Solid State Drive}
}

\newglossaryentry{hdd}
{
    name=HDD,
    description={Hard Disk Drive}
}

\newglossaryentry{scm}
{
    name=SCM,
    description={Storage Class Memory}
}

\newglossaryentry{mmu}
{
    name=MMU,
    description={Memory Management Unit}
}

\newglossaryentry{vm}
{
    name=VM,
    description={Virtual Memory}
}

\newglossaryentry{vma}
{
    name=VMA,
    description={Virtual Memory Area}
}

\newglossaryentry{vfs}
{
    name=VFS,
    description={Virtual File System}
}

\newglossaryentry{aslr}
{
    name=ASLR,
    description={Address Space Layout Randomization}
}

\newglossaryentry{tlb}
{
    name=TLB,
    description={Translation Lookaside Buffer}
}

\newglossaryentry{posix}
{
    name=POSIX,
    description={Portable Operating System Interface}
}

\newglossaryentry{tfs}
{
    name=TFS,
    description={Trusted File System Service}
}

\newglossaryentry{rpc}
{
    name=RPC,
    description={Remote Procedure Call}
}

\newglossaryentry{nvmm}
{
    name=NVMM,
    description={Non-Volatile Main Memory}
}

\newglossaryentry{nvdimm}
{
    name=NVDIMM,
    description={Non-Volatile DIMM}
}

\newglossaryentry{cpu}
{
    name=CPU,
    description={Central Processing Unit}
}

\newglossaryentry{pd}
{
    name=PD,
    description={Persistence Domain}
}

\newglossaryentry{adr}
{
    name=ADR,
    description={Asynchronous DRAM Refresh}
}

\newglossaryentry{eadr}
{
    name=eADR,
    description={Extended Asynchronous DRAM Refresh}
}

\newglossaryentry{wpq}
{
    name=WPQ,
    description={Write Pending Queue}
}

\newglossaryentry{ait}
{
    name=AIT,
    description={Address Translation Table}
}

\newglossaryentry{cow}
{
    name=CoW,
    description={Copy-on-Write}
}

\newglossaryentry{wal}
{
    name=WAL,
    description={Write-Ahead Log}
}

\newglossaryentry{zns}{
    name=ZNS,
    description={Zoned Namespace}
}

\newglossaryentry{woc}{
    name=WoC,
    description={Write-optimized Compressed Key}
}

\newglossaryentry{ppt}{
    name=PPT,
    description={Persistent Page Table}
}

\newglossaryentry{pte}{
    name=PTE,
    description={Page Table Entry}
}

\newglossaryentry{simd}{
    name=SIMD,
    description={Single Instruction Multiple Data}
}

\newglossaryentry{lsm}{
    name=LSM,
    description={Log-Structured Merge}
}

\newglossaryentry{cas}{
    name=CAS,
    description={Compare-and-Swap Operation}
}

\newglossaryentry{cxl}{
    name=CXL,
    description={Compute Express Link}
}

\newglossaryentry{wa}{
    name=WA,
    description={Write Amplification}
}

\begin{document}

\date{}

\providecommand{\keywords}[1]{\textbf{Keywords.} #1}

\title{\Large \bf Persistent Memory File Systems:\\
  A Survey}

\author{
{\rm Wiebe van Breukelen}\\
Vrije Universiteit Amsterdam
\and
{\rm Animesh Trivedi}\\
Vrije Universiteit Amsterdam
} 

\maketitle

\begin{abstract}

Persistent Memory (\gls{pm}) is non-volatile byte-addressable memory that offers read and write latencies in the order of magnitude smaller than flash storage, such as SSDs. This survey discusses how file systems address the most prominent challenges in the implementation of file systems for Persistent Memory. First, we discuss how the properties of Persistent Memory change file system design. Second, we discuss work that aims to optimize small file I/O and the associated metadata resolution. Third, we address how existing Persistent Memory file systems achieve (meta) data persistence and consistency.

\end{abstract}


\section{Introduction} \label{sec:introduction}
Over the past several decades, data storage has become an indispensable part of modern society. However, modern storage had its origins in the early twentieth century. Charles Babbage, who is considered by some to be the "father of the computer", introduced a simplistic form of storage in his Analytical Engine: a general-purpose computer that could be programmed by punch cards~\cite{copeland-modern-2020}. With the emergence of faster and more advanced computers in the 1960s, storage demand grew exponentially. As a result, \emph{magnetic storage}, where data is stored on rotating platters, like on a hard disk drive (\gls{hdd}), quickly gained traction. Until now, this growth has not slowed down.
\par
As storage demands and processing power increased, a new bottleneck emerged. In demanding environments, such as data centers, data access time could not keep up with~\gls{cpu} speed. This speed gap between CPU and storage continues to grow, so faster storage devices are necessary~\cite{fontana-moores-2018, fingerhut-does-2014}. 
\par
A Solid-State Drive (\gls{ssd}), a form of \emph{flash storage}, offers lower read and write latencies than an HDD, especially in a workload that involves a lot of random data accesses~\cite{leventhal-flash-2008}. Like HDDs, SSDs exchange data by the smallest unit of access: a block~\cite{yang-i-cash-2011}. Exchanging these blocks between the computer (or host) and storage efficiently is an ongoing challenge. Compared to CPUs, storage devices are an order of magnitude slower in terms of latency~\cite{gugnani-understanding-2021}. 
\par
Operating systems strive to minimize the impact of high device latency on application speed. For example, the Linux kernel reduces the performance impact as much as possible by maintaining a \emph{page cache}: a chunk of memory where the OS caches chunks of a file for later use. Based on access patterns, disk blocks can be loaded into memory proactively, allowing substantially lower access latencies~\cite{chen-enabling-2018}.
\par
Such mitigations are due to the view we had on storage over the past 50 years. We assumed a two-level storage hierarchy: a fast primary memory (e.g., DRAM) and slow secondary memory (e.g., HDD). Both memories have their own unique properties, for example, the access interface, location within the computer architecture, and access latencies. This has a large influence on the overall design of the Operating System. An alternative scheme, the \emph{one-level storage hierarchy}, changes how we view storage as a whole. Instead of a hierarchy in which we combine the strengths of multiple storage devices, we switch to a hierarchy in which we combine storage and memory into a single device.
Persistent Memory (PM) enables the use of such hierarchy~\cite{bailey-operating-2011}. It is a form of storage that is very related to DRAM in terms of access latency, the most significant difference being that PM is non-volatile while DRAM is volatile. A well-known example of Persistent Memory is Intel's Optane Memory~\cite{izraelevitz-basic-2019}.


\par
\begin{figure}[h]
    \includegraphics[width=1.0\linewidth]{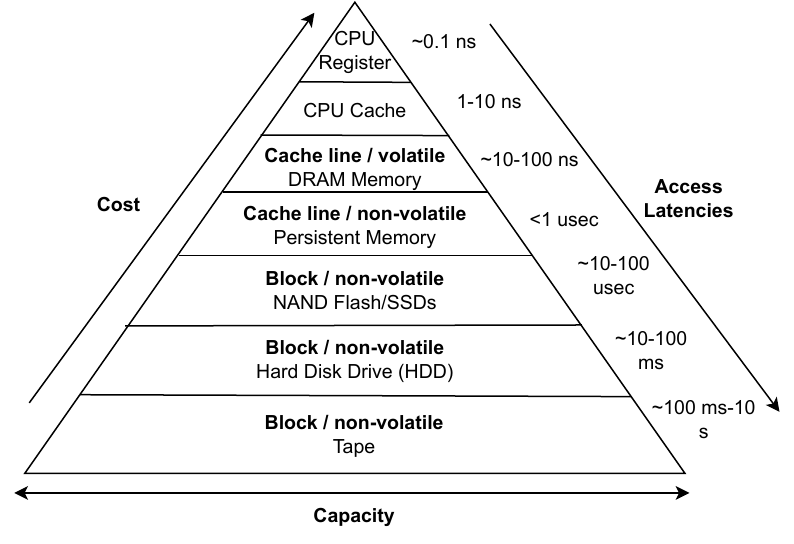}
    \caption{Overview of storage devices and their position within the \emph{Modern Triangle of Storage Hierarchy}}
    \label{fig:triangle-storage}
\end{figure}

To better illustrate the position of PM in the storage hierarchy, consider~\autoref{fig:triangle-storage}. PM is located between an SSD and a DRAM module in terms of access latencies and is accessed through CPU load and store instructions at cache line granularity; $64$ bytes for the \texttt{x86-64} architecture~\cite{waymann-survey-2017}. Note that the capacity and cost scale with the access latencies; storage located at the top (e.g., CPU caches) of the pyramid is scarce and costly compared to storage at the bottom of the pyramid, e.g. HDDs.
In terms of data bandwidth, DRAM outperforms PM by quite a margin, see~\autoref{tab:pm-speedups-overview}.

\begin{table}[h]
    \centering
    \small
    \begin{tabular}{l|c|c|c}
             & \bf DRAM & \bf Intel Optane PM & \bf $\%$ slower \\
        \hline
        Read Bandwidth & 30 GB/s & 7 GB/s & $76.6 \%$\\
        Write Bandwidth & 20 GB/s & $2.2$ GB/s & $89 \%$ \\
    \end{tabular}
    \caption{Intel Optane $256$ GB single DIMM bandwidth ($4$ threads) \texttt{ntstore}-based benchmark performed by Yang et. al~\cite{yang-empirical-nodate}}
    \label{tab:pm-speedups-overview}
\end{table}

\par
In addition to lower latencies, the byte-addressable property of persistent memory introduces even more opportunities. Byte-addressable storage has been around for decades, for instance, in BIOS chips. Persistent memory introduces this concept to storage, allowing applications to define their own persistent data structures in memory. Another advantage is rapid recovery: a rollback can be performed, even if the application is not aware that the memory is non-volatile. Application-critical data structures are transparently made persistent, so no expensive backup/snapshot run is required~\cite{zhang-study-2015}. The emergence of PM brings exciting new possibilities but also new challenges:

\begin{inparaenum}[(1)]
    \item Traditionally, file blocks are accessed through a system call. Within the kernel, this request is passed to the Virtual File System (\gls{vfs}), which performs a metadata lookup, resolves the physical location of the file on the disk or page cache, and finally copies the data into a user-allocated buffer.\par
    In the case of slow block devices, the overhead of trapping into the kernel is negligible, as the latency of the device is much higher than the time of the kernel trap. However, this does not apply to PM, as it can perform storage operations in which the latencies are magnitudes smaller than a kernel trap. Consequently, the overhead shifts from the device to the host I/O stack~\cite{li-ctfs-2022}. \par
    \item A closely related challenge is \emph{indexing overhead}. For example, suppose that an application wants to change the permissions for the file \texttt{/foo/bar/a.txt}, located on a mounted $ext4$ file system. It performs a \texttt{chmod()} call. This call is then forwarded to the kernel's VFS, which in turn performs a so-called \emph{file path walk} by walking the directory tree to locate \emph{file metadata}~\cite{cai-flatfs-2022}.  Then, a metadata operation is performed; in this case, a file permission change.\par
    The major overhead in this example boils down to this path walk. The total time spent is linear to the number of subdirectories in the file path, in the aforementioned example, two. In the case of slow block storage devices, I/O latency dominates, so this delay is acceptable. On the contrary, PM latencies are an order of magnitude smaller and, therefore, such software overhead is unacceptable~\cite{cai-flatfs-2022, koo-modernizing-2021}.\par
    \item The third challenge is related to the persistence of the data. Modern operating systems use hardware and software caches for performance benefits. Recall that in PM, data is accessed using CPU load and store instructions, like DRAM. In the event of a system crash, this may cause data loss, as the data in the caches may not yet be flushed to disk~\cite{bhandari-implications-2012}. A closely related issue is write ordering. If write ordering is enforced, a rollback is trivial, as storage transactions are temporally ordered. However, the Memory Controller, also known as the Memory Management Unit (\gls{mmu}), can bypass this constraint, as it can violate the order of writing in favor of performance~\cite{wu-scmfs-2013}.
\end{inparaenum}

This survey discusses the changes proposed to solve these challenges. First, we cover the design choices and data structures of existing PM file systems at a high level, i.e., the differences in various file system designs, their position within the OS ecosystem, and their respective strengths and weaknesses. \par
Second, we discuss work that aims to optimize small file I/O and the associated metadata resolution. Finally, we discuss how PM file systems achieve data crash consistency.



\section{Study Design}
This section discusses the setup of the literature study. First, we define the survey goal, resulting in one research question and three sub-questions. Subsequently, the scope of the research is defined and, finally, the methodology used for the selection of related literature.

\subsection{Research Goal}

As mentioned in~\autoref{sec:introduction}, the goal of this survey is to investigate the changes required to deal with the three challenges in persistent memory, namely: OS overhead, indexing overhead, and data persistence. Therefore, the main research question is:
\emph{Which file system design changes are needed to cope with the challenges that arise when using Persistent Memory?}
\par

In order to answer this question, we have defined three sub-questions. Each sub-question covers one of the three challenges relevant when designing persistent memory file systems: 

\begin{questions}[itemindent=1em]
    \item How have the properties/features of Persistent Memory led to changes in file system design?~\label{qs:fs-design}
    \item Which optimizations help to decrease file indexing overhead in small I/O workloads? \label{qs:metadata-perf}
    \item How can Persistent Memory file systems guarantee data crash consistency?  \label{qs:persistency}
\end{questions}

\subsection{Scope}
In this study, we focus on the challenges in implementing file systems for Persistent Memory. Papers in related fields, such as Persistent Key-Value (\gls{kv}) stores/databases, are \emph{only} included if they explicitly aim to exploit PM advantages for better FS performance. Note that this assessment is based on the impression of the paper's abstract and conclusion, unless stated otherwise.\par
We can illustrate this with an example. RocksDB, a persistent KV store, uses the advantages of NOR or NAND Flash \gls{ssd}s to store associative arrays efficiently. It achieves significant space efficiency and better write throughput while achieving acceptable read performance~\cite{debnath-flashstore-2010, dong-rocksdb-2021}.\par
Recent work has shown that RocksDB can be adapted to work with PM, although this implementation is very experimental and poorly documented~\cite{noauthor-rocksdb-nodate}. In such cases, we opt to exclude RocksDB from the literature study.\par
In conclusion, we limit this survey to the available literature that \emph{specifically} aims to address the challenges, opportunities, or future work when implementing \gls{pm}-based file systems.

\subsection{Methodology}
In this survey, we use the \emph{Snowball} sampling~\cite{wohlin-guidelines-2014} methodology to find relevant papers. In this search methodology, one starts by reading the so-called \emph{seed papers}. These papers are then used to find other relevant articles. The seed papers used for this study are listed in~\autoref{tab:seed-papers}. In order for a paper to be selected, it must adhere to the following inclusion criteria, and none of the exclusion criteria:

\begin{itemize}
    \item I.1 - The work proposes a new file system, or is a continuation/improvement of an existing file system;
    \item I.2 - The work especially targets Persistent Memory or related terms: Storage Class Memory (SCM), Phase Change Memory (PCM), or Non-Volatile Main Memory (NVM/NVMM). One or more of these terms must be present in the paper's abstract;
    \item I.3 - The work is released after 2005. We selected this cut-off date because this year Intel announced PCM modules that could replace traditional flash cards~\cite{spooner-intel-2006};.
    \item E.1 - The work proposes a new data storage paradigm built on top of an existing file system, for example, a key-value store or hash table. Papers that propose using such paradigms to improve file systems are exempted from this criteria;
    \item E.2 - The work is classified as a literature study;
\end{itemize}

Note that in rare cases, we might violate constraints, e.g., when discussing relevant background theory.
\par
In addition to Snowball Sampling, we also performed a manual search. The most relevant keywords are listed in~\autoref{tab:search-keywords}. We prefer work that is well-cited (i.e, $>20$ citations), builds on well-established papers, and is published in the last five years; 2017 - 2022. The names of the conferences considered are: \emph{USENIX (FAST)}, \emph{EuroSys}, \emph{ODSI}, \emph{ASPLOS}, \emph{ACM} (\emph{SIGARCH}, \emph{SIGOPS}, \emph{SPAA}), \emph{IEEE}, \emph{SOSP}, \emph{VLDB}, \emph{SC}, and \emph{ISCA}. In addition, we use \emph{ACM}, \emph{Arxiv}, and \emph{Google Scholar} search engines to find additional work and background theory.

\begin{table}[h]
    \centering
    \small
    \begin{tabularx}{0.5\linewidth}{ p{0.3\linewidth} p{0.1\linewidth}}\toprule[1.5pt]
    \bf Title & \bf Year\\\midrule
    BPFS~\cite{condit-better-2009} & 2009\\
    SCMFS~\cite{wu-scmfs-2013} & 2013\\
    Aerie~\cite{volos-aerie-2014} & 2014\\
    HiNFS~\cite{ou-hinfs-2016}      & 2016\\
    Strata~\cite{kwon-strata-2017} & 2017\\
    NOVA~\cite{xu-nova-2016} & 2017\\
    SplitFS~\cite{kadekodi-splitfs-2019} & 2019\\
    HashFS~\cite{neal-hashfs-2021} & 2021\\
    Kuco~\cite{chen-kuco-2021}  & 2021\\
    FlatFS~\cite{cai-flatfs-2022} & 2022\\
    \bottomrule[1.25pt]
    \end {tabularx}
    \captionof{table}{Seed Papers, shortened titles} \label{tab:seed-papers} 
\end{table}

\begin{table}[h]
    \centering
    \small
    \begin{tabularx}{1.0\linewidth}{ l c c }\toprule[1.5pt]
    \bf Keyword & \bf Accepted & \bf Rejected\\\midrule
    Persistent Memory File System  & 12 & 6 \\
    NVM(M) File System & 1 & 35 \\
    Intel Optane File System & 3 & 6 \\
    Direct Access/DAX File System & 2 & 2 \\
    Persistent Memory Kernel Bypass & 0 & 8\\
    \bottomrule[1.25pt]
    \end {tabularx}
    \captionof{table}{Exploratory Keywords and Paper inclusion/exclusion count} \label{tab:search-keywords} 
\end{table}

\section{Background} \label{sec:background}

Before going into the changes required to build fast PM file systems, we first provide the relevant background theory. First, ~\autoref{sec:background-hw} discusses the relevant hardware concepts depicted in~\autoref{fig:pm-arch-hw}. Subsequently, \autoref{sec:background-sw} shifts the focus to the software components within the kernel, using~\autoref{fig:pm-arch-sw} as a guideline. Finally, \autoref{sec:data-structures} discusses well-established data structures used in modern file systems.

\subsection{Persistent Memory Hardware Architecture} \label{sec:background-hw}

\begin{figure}[h]
    \centering
    \includegraphics[width=0.9\linewidth]{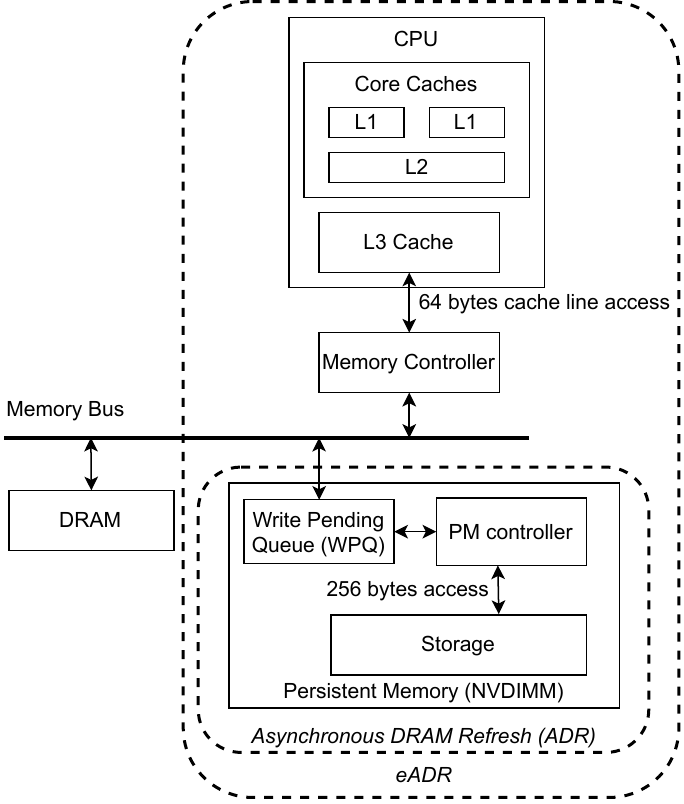}
    \caption{Position of Persistent Memory within the CPU hardware architecture}
    \label{fig:pm-arch-hw}
\end{figure}

As mentioned in the introduction, persistent memories are byte-addressable. Therefore, a CPU can issue read/write requests using similar microprocessor memory instructions issued when accessing DRAM. \autoref{fig:pm-arch-hw} displays the position of PM within the computer hardware architecture. When the CPU issues a read request, it requests memory at a particular memory address in DRAM/PM. Now, we have two options: the data is in the cache and can be immediately transferred to the CPU (\emph{cache hit}, faster), or the data should be fetched from DRAM/PM using the \emph{memory controller} (\emph{cache miss}, slower). 
\par

\emph{CPU cache}:
Modern CPUs use a hierarchical caching architecture consisting of `levels' to reduce the access latency to main memory~\cite{su-survey-2021}. For example, for the \texttt{x86-64} architecture, the first cache level (L1) offers lower latencies than a level 2 (L2) or level 3 (L3) cache.  On the contrary, an L3 cache offers greater capacity than an L2 cache. For instance, CPUs based on Intel's \emph{Ice Lake} architecture have $48$ kB of L1 cache, $512$ kB of L2 cache, and $6$ MB L3 cache. In terms of latency, an L1 cache line fetch takes approximately $5$ CPU cycles, while an L3 cache fetch costs $\sim 41$ cycles~\footnote{In case of a cache hit. If requested data do not reside in the cache (i.e., a \emph{cache miss}), the access latency increases as the data must be fetched from main 
memory~\cite{su-survey-2021}.}~\cite{saarland-informatics-campus-cache-nodate}.  
\par
In addition to latency and capacity, three properties are especially important when considering PM~\cite{dulloor-pmfs-2014, wu-scmfs-2013, bhandari-implications-2012, lu-loose-ordering-2014}. First of all, data resident in the cache may be shared between cores, allowing better multiprocessing performance. Second, in contrast to PM, the cache may not be byte-addressable. In the case of the \texttt{x86-64} architecture, the cache stores data in fixed fragments of $64$ bytes, also known as a \emph{cache line}. As this is the smallest unit of access, the data is fetched or written in $64$ byte chunks.\par
\begin{figure}[h]
    \centering
    \includegraphics[width=0.9\linewidth]{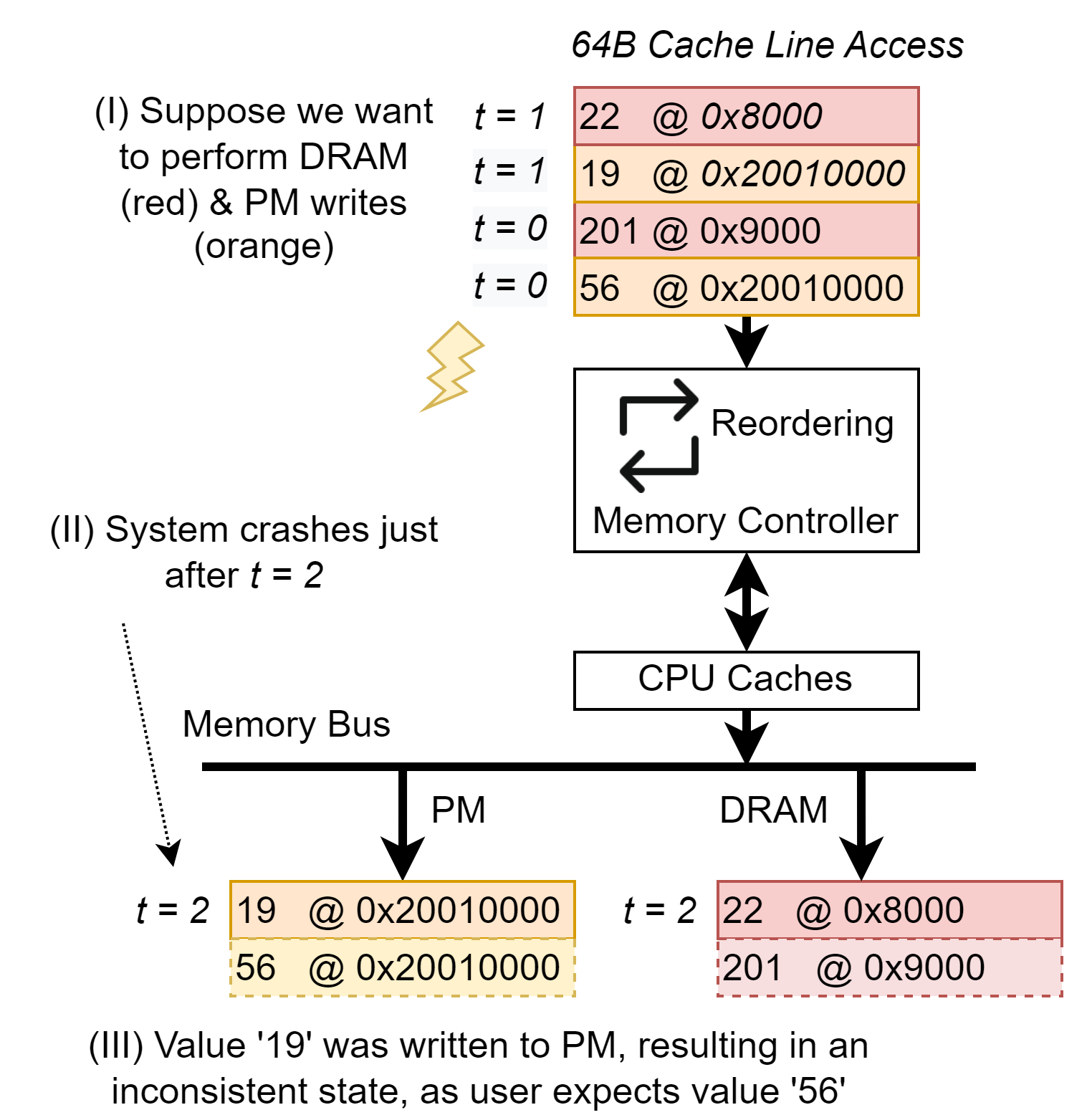}
    \caption{Implications of Write Reordering for Persistent Memory}
    \label{fig:cache-reordering-implications}
\end{figure}

Third, in shared memory systems, where memory is shared between multiple cores, caches need to confirm to a \emph{cache coherence protocol}. This protocol ensures that each core always has access to the most up-to-date version of a memory location contained in the cache~\cite{caheny-reducing-2018}. However, adhering to such a protocol at the price of performance, as it requires additional CPU cycles to perform the necessary consistency checks. To amortize this impact, CPUs may reorder writes~\cite{mckenney-memory-2005}. Although this might be beneficial for DRAM performance, it has a big implication for PM which we demonstrate by the example provided in~\autoref{fig:cache-reordering-implications}.
\par

In this example, the CPU memory controller is instructed to perform four DRAM/PM writes in total. For instance, at $t=0$ a file system commits the value `$56$' to address \texttt{0x20010000} in PM and the value `$201$' to DRAM. Note that these requests are not processed immediately; instead, they are buffered at the CPU memory controller. At $t=2$, the memory controller decides to delay the writes at $t=0$ in favor of two other writes posted at $t=1$. Just after processing these two writes, the system crashes. Consequently, both the PM and DRAM writes at $t=0$ are lost forever. Due to the non-volatile property of PM, the file system ends up in an inconsistent state. The file system (falsely) assumes that file changes are committed to PM in chronological order~\cite{lu-loose-ordering-2014}, therefore, it assumes that the value $56$ is present at the location \texttt{0x20010000}. Instead, this location now contains the value `$19$' due to the aforementioned write reordering.

\par
Summarizing, write reordering may be beneficial for DRAM reads/writes; however, it poses a threat to PM transactions~\cite{wu-scmfs-2013, volos-aerie-2014, li-ctfs-2022, yang-spmfs-2021}. First of all, the order data is being written to PM may differ from the user's intentions. Second, in case of a system crash, the data could still be located in the cache and not yet flushed to PM~\cite{bhandari-implications-2012}. For now, it is important to understand the implications of CPU caches for PM. The available solutions to fix this issue will be discussed in~\autoref{sec:persistency-atomicity}.
\par
\emph{Memory controller}:
In the case of a \emph{cache miss}, the data must first be obtained from DRAM; therefore, the request is sent to \emph{memory controller}, also known as a Memory Management Unit (\gls{mmu}). An MMU serves two purposes: translation and isolation~\cite{wu-scmfs-2013}. Translation comes down to converting virtual addresses into physical (device) addresses. An MMU provides memory isolation for security and efficient memory fragmentation between applications and the kernel. 
The internals of an MMU are quite complex; therefore, we refer to the Linux kernel documentation~\cite{linux-foundation-page-nodate}, which we quickly summarize.
\par
An MMU uses a \emph{page table} to map virtual addresses to physical addresses. The \emph{x86-64} architecture uses a hierarchical structure in which virtual addresses have a length of $48$ bits. The $16$ MSB bits cannot be used. This tree structure consists of four levels~\footnote{Since the Intel Ice Lake microarchitecture (released in 2017) the CPU supports five-level page tables, extending the size of a virtual address to $56$ bits, resulting in 128 petabyte addressable space~\cite{noauthor-intel-2022}.}, where each level can accommodate for $512$ mappings. At level 3, each mapping covers one $4$ kB page, so in total $2$ MiB of physical memory as $4 \text{kiB} * 1024 = 2 \text{MiB}$, that is, the last 12 bits of a virtual address. At Level 2, each mapping covers one $2$ MB huge page, i.e., the last 19 bits of the virtual addresses~\cite{iaik-paging-nodate}. A key observation here is that when the level decreases, the associated mappings become coarse-grained, therefore, mapping larger regions of memory.
\par

\par
\emph{Persistence Domain}:
Another important concept is the Persistence Domain (\gls{pd}). As mentioned before, PM is non-volatile memory. Therefore, if a crash occurs, persistence must be preserved, unlike DRAM, where data loss is inevitable. A persistent domain is a region within a computer system where data is guaranteed to be persistent, even in the event of system failure~\cite{noauthor-nvm-2017}. \autoref{fig:pm-arch-hw} depicts two persistence domains, namely, Asynchronous DRAM Refresh (\gls{adr}) and enhanced ADR (\gls{eadr}). The first guarantees data persistence within the PM device, while the second extends this guarantee to the CPU cache~\cite{rudoff-deprecating-2016}.
\par
We now switch our focus to the internals of the PM device depicted at the bottom right of~\autoref{fig:pm-arch-hw}. Device commands end up in a buffer queue, the Writing Pending Queue (\gls{wpq}). This queue is part of both the ADR and eADR domains. In case of a crash or power failure, this buffer is emptied and written to storage using the remaining electrical charge in a (super) capacitor.\par
After passing the WPQ, the writes end up at the PM controller. This controller includes an Address Translation Table (\gls{ait}), which maps physical addresses to device addresses and performs wear leveling. Wear leveling prolongs the lifetime of PM by spreading the Program and Erase (P/E) cycles over the memory cells. This is important because flash cells can only endure a finite number of cycles~\cite{izraelevitz-basic-2019}.
\par
Eventually, the data is written to PM in fixed-sized chunks, in the case of Intel Optane $256$ bytes. As mentioned above, the CPU accesses DRAM via 64 $bytes$ load and store instructions. Consequently, writing less than $256$ bytes will result in \emph{write amplification}: the difference between the actual amount written and the amount of data intended to be written~\cite{peng-system-2019}.
\par
We can conclude that PM devices have interesting hardware properties, especially issues related to write reordering and the persistence domain. Both properties are fundamental in evaluating the design of PM file systems.

\subsection{Storage Software Stack} \label{sec:background-sw}
In order to utilize storage devices efficiently, software is essential. 
This subsection discusses software aspects relevant when working with storage devices. More specifically, we discuss the two most prominent forms of accessing storage from user space: conventional system calls and Direct Address (\gls{dax}) via the MMU. These techniques can be considered `building blocks' for more advanced PM file system designs which we discuss later.
\par

\begin{figure}[h]
    \centering
    \includegraphics[width=1.0\linewidth]{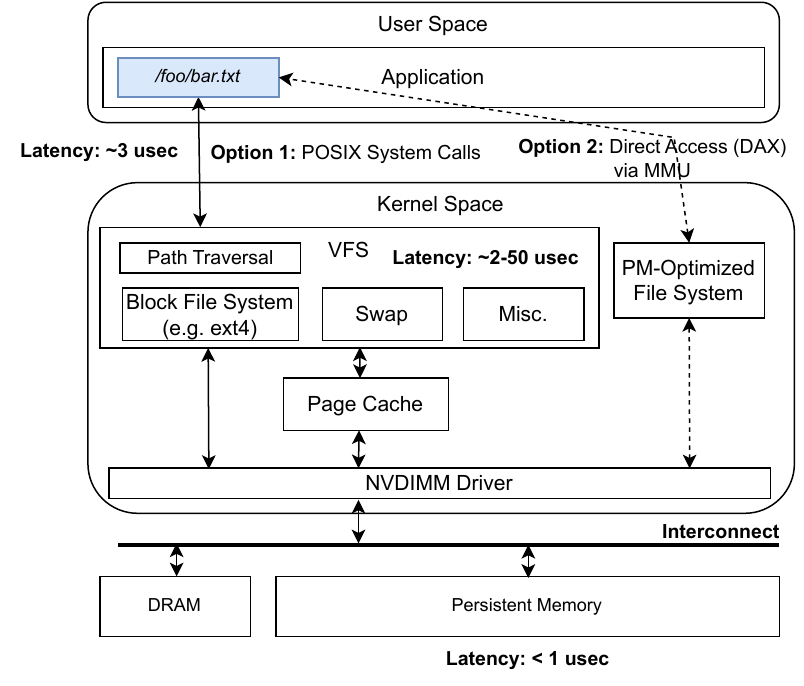}
    \caption{Software Architecture Persistent Memory}
    \label{fig:pm-arch-sw}
\end{figure}

\autoref{fig:pm-arch-sw} depicts the position of PM in the software stack. We briefly elaborate on the most important components.
\par
In conventional two-level file systems, applications in \emph{user space} request FS-related services through \emph{system calls}. For example, the \texttt{open()} system call can be used to open a file handle, \texttt{write()} to write to a file, and \texttt{close()} to close a file handle. In kernel space, these FS-related system calls are directed to the Virtual File System (\gls{vfs}). In order to support multiple file systems, the VFS implements a generic software interface that is independent of the actual file system, e.g.: \texttt{ext4}, swap space, or a PM-optimized file system. Internally, the VFS performs a \emph{path traversal}, which recursively traverses the file path, for example, "\texttt{/foo/bar.txt}", and informs the corresponding file system of an incoming request. Now, the file system looks up the corresponding \emph{file metadata} and performs the requested action~\cite{cai-flatfs-2022}. 
\par
As mentioned before, the Linux Kernel uses a page cache to decrease the performance impact of slow device latencies. As a result, file data may (partly) reside on the storage device or in the \emph{page cache}:\par
\begin{inparaenum}[(1)]
    \item The file is (partially) located in the page cache. In this case, the requested data can be returned immediately at a very low cost. \par
    \item The requested file blocks are not in the cache. A request is sent to the corresponding device driver, which in turn sends the actual command to the storage device.\par
\end{inparaenum}

The use of a page cache in the case of PM is controversial, as the induced latency of the VFS is higher than the access latency of PM~\cite{li-ctfs-2022, wu-scmfs-2013, condit-better-2009, chen-kuco-2021}, as seen in~\autoref{fig:pm-arch-sw}. 
\par
An alternative technique to access file data is Direct Access (\gls{dax}). Using DAX, pages are directly mapped in user space using the MMU, avoiding the performance-degrading VFS and page cache~\cite{linux-foundation-direct-nodate}.  Although this concept is very appealing, it has one (potential) disadvantage: kernel and file system guarantees are lost as one can access storage without kernel interference. 
\par
In short, there are two prominent methods to access storage: System Calls and DAX. DAX avoids the use of a page cache. 

\subsection{Traditional File System Structures} \label{sec:data-structures}
This subsection covers the most prominent data structures found in file systems: metadata structures, B+ trees, LSM trees, and extent/radix trees.

\paragraph{Metadata structures}
File systems rely on \emph{metadata} to gain insight into files and directories stored within the file system. It is stored on the storage device alongside the actual data blocks. The most important use of metadata is to enable \emph{file mapping}:  the operation of mapping a logical file offset to a physical location on the underlying storage device~\cite{neal-hashfs-2021}. Generally, file systems map files at block granularity, in most cases 4 $kB$, to constrain the amount of metadata required.\par
File systems can define any metadata structure to better accommodate their requirements. For example, the \emph{ext2} file system uses \emph{inodes}. An inode is a per-file structure~\footnote{In \emph{ext2}, a directory is considered a special type of file.} that contains elementary data fields such as file creation modification date, file size, and permissions. Inodes are stored in a \emph{inode table} on disk. To accommodate file mapping for a wide variety of file sizes, \emph{ext2} uses a scalable structure, more specifically an \emph{extent tree}, which we will discuss later. Nodes within this tree represent logical to physical block mappings.

\paragraph{B+ tree}

A B tree is a self-balancing tree, which means that the node keys are sorted in ascending order, enabling fast sequential performance~\cite{comer-ubiquitous-1979}. Compared to a B tree, a B+ tree only stores values at the bottom of the tree using linked leaves. The internal (non-leaf) nodes only contain keys~\cite{chen-persistent-2015}.

\begin{figure}[h]
    \centering
    \includegraphics[width=1.0\linewidth]{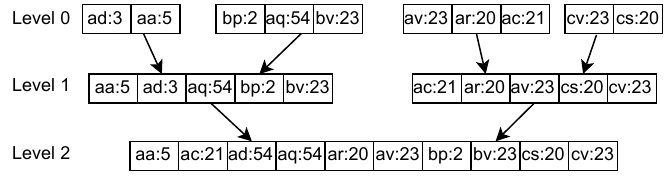}
    \caption{Three-level Log-Structured Merge (LSM) tree}
    \label{fig:lsm-tree}
\end{figure}

\paragraph{LSM Tree}
A Log-Structured Merge (\gls{lsm}) Tree is a disk-optimized search tree for storing Key-Value (\gls{kv}) pairs~\cite{oneil-log-structured-1996}. \autoref{fig:lsm-tree} depicts a three-level LSM tree. Observe that in this figure, level 0 is an unsorted append-only log located in DRAM. If a level 0 log runs out of space, a compaction routine is started in the background. This compact routine performs Garbage Collection, which in essence iteratively compacts KV pairs at level $x$ into larger, sorted segments located at level $x+1$. Subsequently, the segments are flushed to disk.\par
An LSM tree avoids Write Amplification (\gls{wa}) by performing sorting in the background; users can write directly to the in-DRAM log.

\begin{figure}[h]
    \centering
    \includegraphics[width=1.0\linewidth]{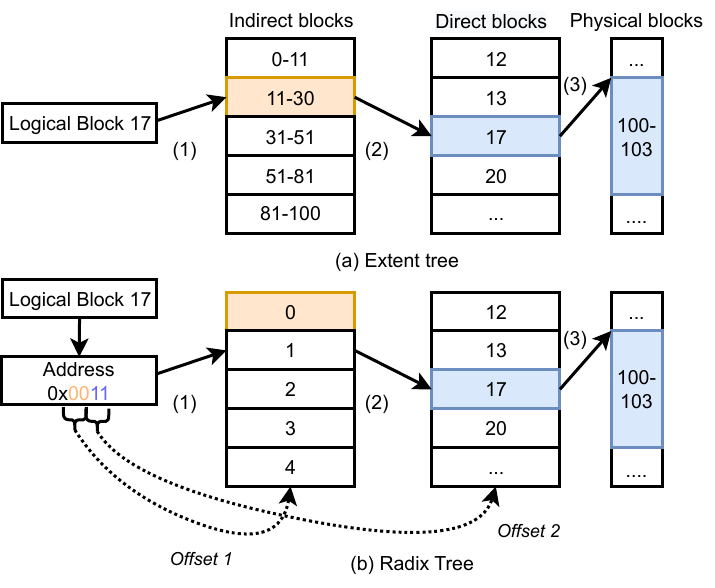}
    \caption{Example of Extent Tree (a) and Radix Tree (b) logical to physical block translation}
    \label{fig:extent-and-radix-tree}
\end{figure}

\paragraph{Extent and Radix Tree}
Both the Extent and Radix tree are B-trees. An \emph{extent} is a data structure that represents a range of contiguous physical blocks, e.g., $11-30$ in~\autoref{fig:extent-and-radix-tree}a. Together, the extents can form a tree that allows efficient \emph{logical block} to physical block number translation. Compared to \emph{file offset}, the offset from the beginning of a file, a logical block number is defined in the OS as a multiple of the device \emph{block size}, usually $512$ bytes. The physical block number represents this location on the actual storage device~\cite{mathur-new-2007}. In \autoref{fig:extent-and-radix-tree}, the logical block $17$ is translated into the physical block $100$. Indirect blocks are included to enable file growth and shrinking operations.
\par
Radix trees use a different mapping scheme. Instead of using the block number, it uses the corresponding binary representation to perform the lookup. For example, we can map every group of 9 bits to one Radix node, as shown by the dotted arrows in~\autoref{fig:extent-and-radix-tree}b.\
\par
An advantage of Extent trees is that they consume less memory than Radix trees, as they grow slower over time. However, Radix trees are computationally less expensive compared to Extent trees as they only require simple arithmetic offset operations for lookups~\cite{neal-hashfs-2021}.

\subsection{Crash-Consistency Techniques} \label{sec:crash-consistency-techq}
Crash consistency guarantees are essential for file systems. Without these guarantees, a crash may cause data corruption or, in extreme cases, leave the file system inoperable. Some associate data persistence exclusively with the actual data blocks stored on the device; however, data persistence must be enforced in multiple areas ~\cite{xu-nova-2016}. For example, writing to a file involves updating the corresponding data blocks, the last modification date, and the file length. Therefore, in addition to data consistency, metadata consistency is essential.\par
Generally speaking, modern crash consistency techniques can be categorized into three areas: \emph{Journaling}, \emph{Shadow Paging}, and \emph{Log-Structuring}~\cite{dulloor-pmfs-2014, condit-better-2009, xu-nova-2016}. We briefly discuss these techniques below.

\begin{figure}[h]
    \centering
    \includegraphics[width=0.8\linewidth]{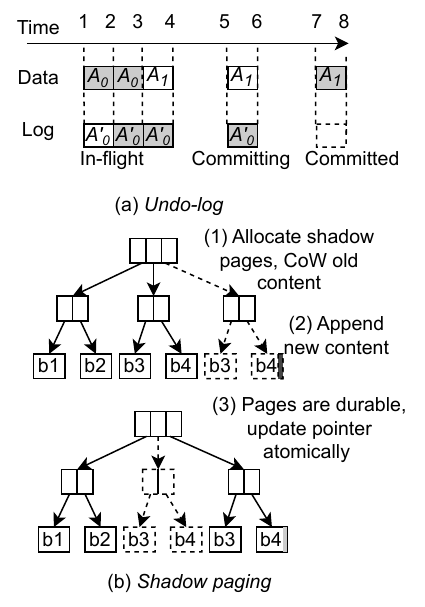}
    \caption{Crash-consistency techniques, (a): \emph{undo-log}, (b): \emph{shadow-paging}}
    \label{fig:crash-consistency-techniques}
\end{figure}

\paragraph{Journaling}
A journal is a data structure that keeps track of changes in the file system, separated from the on-disk data blocks. In essence, it is a chronologically ordered log of (meta)data changes, i.e., transactions. A transaction consists of operations that are \emph{idempotent}, which means that they can be repeated infinitely many times without side effects~\cite{wan-empirical-2016}. If a crash occurs, the file system can be restored to a consistent state using a Write-Ahead Log (\gls{wal}): an append-only disk-resident structure used for crash recovery. In journaling, there are two types of WAL, namely \emph{undo logging} and \emph{redo logging}~\cite{wan-empirical-2016}. 
\par
In undo logging, a copy of the original data is inserted into the log before a transaction starts. In the event of a rollback, the modifications are restored using the data contained in the log. This process is depicted in~\autoref{fig:crash-consistency-techniques}a. In this example, events are temporally ordered, just as in the journal log. At $t=1$, the transaction starts by pushing the original data to the log: $A'-0$. At $t=2$, this data is in stable storage, indicated by the grey coloring. In-flight data can now be sent to the storage device. When the transaction eventually \emph{commits} ($t=5$) and the changes are in stable storage, the original data can be erased from the log ($t=7$).
\par
In redo logging, transactions append data modifications to the log. Only when the transaction commits the corresponding data changes are stored in PM.
\par
Undo/redo logging has its own (dis)advantages. The most crucial difference observed is that redo logging outperforms undo logging in transactions that update a large number of different objects, while it underperforms in workloads with intensive read operations~\cite{wan-empirical-2016}.
As performance is heavily affected by the type of workload, there is no consensus on which form of logging is best. However, in the PM literature, there appears to be a slight bias towards 'undo logging', as it is easier to implement~\cite{dulloor-pmfs-2014, ou-hinfs-2016}. Alternatively, one can also use an \emph{operation log}: a log that only stores file/directory operations, for example, \texttt{APPEND data, \# num bytes, filename, inode number}~\cite{kwon-strata-2017, chen-kuco-2021, kadekodi-splitfs-2019}.

\paragraph{Shadow Paging}
\emph{Shadow Paging} is a consistency method for file systems based on the \emph{Copy-on-Write} (\gls{cow}) technique: a technique that defers resource duplication to the last possible moment, also known as \emph{lazy copying} or \emph{implicit sharing}~\cite{condit-better-2009}. Operating Systems use CoW to increase \emph{memory paging} performance. Memory paging is a technique to read/write data in the smallest unit of data storage access (usually $512$ bytes), \emph{blocks}, from a storage device for use in main memory~\cite{iaik-paging-nodate}. An OS accesses main memory at \emph{page} granularity, usually $4096$ bytes. In CoW-based memory paging, a page copy initially refers to the original page (to save resources) and is copied at the last moment, i.e., when a write comes in~\cite{noauthor-btrfs-2021, condit-better-2009}.\par
File systems that use Shadow Paging use trees to structure metadata and file blocks, see~\autoref{fig:crash-consistency-techniques}b. In the event of a file modification, the original block content is copied to a new page. Then, file modifications are performed on this copied block. When the transaction completes, the changes become persistent by modifying the pointer to the new pointer block. An example is provided in~\autoref{fig:crash-consistency-techniques}b. In this example, a user writes to the block $b4$, which triggers a CoW for the blocks contained in the same \emph{pointer block}. Data modifications are made by modifying the corresponding copied blocks. Eventually, the transaction commits by changing the pointer value in the pointer block located at the top of the tree~\cite{cai-survey-2021, dulloor-pmfs-2014}.

\paragraph{Log-Structuring}
In contrast to Journal-based file systems, which keep track of changes in a separate log on the disk, log-structured file systems store file system metadata and data updates together. This implies that all file system data is structured in the form of a log, also known as \emph{Log-structured File Systems} (LFSs)~\cite{agarwal-journaling-2020}. Originally, LFSs were designed to improve HDD performance, as HDDs offer poor random performance but high sequential performance~\cite{seltzer-implementation-1993}. Although SSDs offer improved random performance, they still perform better in sequential workloads~\cite{chen-understanding-2009}, therefore, it is still beneficial to use sequential accesses as much as possible.
\par 
In an LFS, random writes are buffered in DRAM and merged into large sequential writes to the log. To avoid fragmentation, a periodic \emph{Garbage Collection} (GC) run is performed, in which the free blocks are coalesced to form new contiguous free regions.

\par
In summary, we have seen three file system crash-consistency techniques: \emph{Journaling}, \emph{Shadow Paging}, and \emph{Log-Structuring}. Journal-based file systems maintain a chronologically ordered log of file transactions, stored separately from the on-disk data blocks. Shadow Paging uses the Copy-on-Write technique to ensure data durability. In Log-Structured file systems, the disk becomes one long log, containing all (meta) data transactions.

\section{Persistent Memory File System Design} \label{sec:file-system-design}
Unlike classic block devices, such as HDDs, which communicate through a relatively slow ACHI controller, PM is placed on the memory bus and accessed via processor load-and-store instructions. This change shifts the main performance bottleneck from device to software.
\par
Traditional file systems like \emph{ext2} perform expensive operations on the critical path, for example, logging file updates, file metadata lookup, maintaining persistency, etc. Furthermore, the hierarchical structure in ext2 introduces a high indexing overhead; in the worst case, up to $45\%$ of a simple $4$ kB data append~\cite[p. 37]{li-ctfs-2022}, and up to $4\times$ write amplification~\cite{mohan-analyzing-2017}. \par

To some extent, this performance impact is mitigated by using the page cache. However, in the case of PM, its byte-addressable properties allow us to access data faster than the I/O stack, making an expensive page cache redundant. Therefore, an efficient design and accompanying data structures are crucial for low-latency/high-throughput PM file systems.\par

Considering the design of the file system at a high level, the relevant literature can be categorized into multiple distinct areas, as shown in~\autoref{tab:fs-designs}. To answer sub-question~\ref{qs:fs-design}, we focus on the challenges each design and its associated file systems intend to solve. 
Specific optimizations (e.g.: guaranteeing data persistence/atomicity, indexing overhead reduction) will be addressed later in Sections~\ref{sec:file-indexing-overhead} and~\ref{sec:persistency-atomicity}, respectively. 

\begin{table*}[h]
    \centering
    \small
    \begin{tabularx}{\linewidth}{p{0.18\linewidth} p{0.10\linewidth} p{0.10\linewidth}  p{0.03\linewidth}  p{0.05\linewidth} p{0.40\linewidth}}\toprule[1.5pt]
    \bf High-level Design & \bf File System & \bf User/Kernel Space \bf & \bf DAX & \bf POSIX-compliant & \bf Main Contribution\\\midrule

      Influenced by Traditional File Systems (e.g. \emph{ext2}) & & & \\
      & BPFS~\cite{condit-better-2009} & Kernel & $\xmark$ & $\checkmark$ & POSIX-compliant file system that reduces write amplification through adapted shadow paging \\
      & PMFS~\cite{dulloor-pmfs-2014} & Kernel & $\checkmark$ & $\checkmark$ & Bypass OS page cache and generic block layer, avoid extensive I/O stack modifications. Lightweight in-place metadata updates\\
      & HiNFS~\cite{ou-hinfs-2016} & Kernel & $\checkmark$ & $\checkmark$ &  Elimination of double copy overhead in kernel\\
      & Ext4-DAX~\cite{linux-foundation-direct-nodate} & Kernel & $\checkmark$ & $\checkmark$ & Include DAX to PM in the existing \emph{ext4} file system\\
     
      Contiguous File Allocation & & & \\
      & SCMFS~\cite{wu-scmfs-2013} & Kernel & $\xmark$ & $\xmark$ & Bypass the generic block layer and perform file mapping via the MMU \\
      & SplitFS~\cite{kadekodi-splitfs-2019} & Hybrid & $\checkmark$ & $\xmark$ & Introduces a \emph{hybrid} architecture in which data operations are handled in user space, while metadata operations are processed in the kernel \\
      & Aerie~\cite{volos-aerie-2014} & Hybrid & $\xmark$ & $\xmark$ & Allow user space applications to update metadata directly in user space \\
      & Kuco~\cite{chen-kuco-2021} & User & $\checkmark$ & $\checkmark$ & Address the poor scalability of existing PM hybrid file systems (e.g., SplitFS) \\
      & ZoFS~\cite{dong-performance-2019} & User & $\checkmark$ & $\checkmark$ & Like Aerie, allow user space applications to update metadata directly in user space, however, with less kernel involvement \\

      Log-Structured & & & \\
      & NOVA~\cite{xu-nova-2016} & Kernel & $\xmark$ & $\checkmark$ & Per inode logs to allow massive parallelism, while providing strong consistency guarantees\\
      & Strata~\cite{kwon-strata-2017} & Hybrid & $\checkmark$ & $\checkmark$ & Capture unique properties of multiple storage devices in one file system \\
        
    \bottomrule[1.25pt]
    \end {tabularx}
    \captionof{table}{PM File Systems categorized by their high-level design} \label{tab:fs-designs} 
\end{table*}

\subsection{Influenced by Traditional File Systems} \label{sec:trad-design}
This category of work adapts well-established data structures in file systems, such as the inode tree, to work with PM. Such file systems benefit from upstream fixes/patches in the Linux Kernel, which can then easily be integrated into the PM-optimized file system.

\paragraph{BPFS} BPFS, released in 2009, is, to the best of our knowledge, the first file system adapted to work with PM. It maintains an indirect block tree similar to conventional file systems, i.e., $ext2$. Its main contribution is a PM-optimized implementation of Shadow Paging (see~\autoref{sec:crash-consistency-techq}), \emph{Short-Circuit Shadow Paging}, which we now elaborate.\par

 Although Shadow Paging is a well-proven feature for ensuring consistency, the authors of the BPFS paper~\cite{condit-better-2009} name a clear disadvantage: \emph{Write Amplification} (\gls{wa}). In tree-based file systems, when new data is written in CoW-fashion, the pointers in the parent blocks must also be updated~\cite{condit-better-2009, dulloor-pmfs-2014}, propagating tree node updates upwards the tree  (see~\autoref{fig:crash-consistency-techniques}b). The resulting WA is significant as the smallest addressable unit of storage is, in most cases, $512$ or $4096$ bytes~\cite{cox-jedec-nodate}, while a pointer update is only $8$ bytes in the case of a $64$-bit system.

BPFS avoids this WA by performing in-place updates within data blocks, taking advantage of PM's unique byte-addressable properties. As mentioned before, BPFS uses a tree structure to store inodes. A significant difference is that BPFS also includes the actual file blocks within this tree, forming one giant tree consisting of inodes at the top and the corresponding file data blocks at the bottom. Now, block changes can be made by performing in-place updates, in-place appends, or partial CoW. In-place updates/appends can be performed in case the data block is located in a tree leaf. Partial CoW is used when multiple (non-leaf) file blocks are affected.
\par
At the time BPFS was released, the availability of PM was very poor. Therefore, the benchmark results should be taken with a grain of salt. Still, compared to Microsoft's NTFS, BFTS achieves significantly higher throughput in small I/O workloads. However, for large I/O workloads (e.g., moving files between directories), this overhead is still substantial~\cite{xu-nova-2016}.

\paragraph{PMFS} PMFS is very similar to BPFS in the sense that both implement a POSIX-compliant file system using system calls. Like PMFS, the authors of BPFS acknowledge that file system consistency imposes a major performance penalty. However, compared to BPFS, there are three differences. First, PMFS uses a B tree instead of the conventional indirect block tree used in BPFS for faster file indexing. Second, PMFS enables applications Direct Access (\gls{dax}, \autoref{sec:background-sw}) to PM memory via a \texttt{mmap} interface, bypassing the expensive OS page cache.
Third, PMFS proposes a hybrid approach to handle file (metadata) consistency. Recall that BPFS uses an optimized version of Shadow Paging to ensure consistency. Although this technique certainly decreased WA, the authors of PMFS~\cite{dulloor-pmfs-2014} found that PBFS fails to capitalize on PM's unique byte-addressable property. They observed that metadata updates are usually small ($\leq 64$ bytes). Therefore, PMFS proposes an alternative methodology that performs writes at cache-line granularity using atomic store instructions~\footnote{Intel's atomic store instructions: Section 8.2.4 - Intel Architecture Software Developer Manual~\cite{noauthor-intel-2022}}. For metadata updates larger than a single cache line, PMFS resorts to more expensive \emph{undo logging} (\autoref{sec:data-structures}). For larger data updates, PMFS falls back to Shadow Paging.
\par
In short, PMFS performs atomic updates when possible, as those are the cheapest operations. Only in cases where in-place updates are not atomic, PMFS resorts to undo logging and Shadow Paging~\cite{dulloor-pmfs-2014}.

\paragraph{HiNFS}
The PM file systems we discussed so far mainly improve performance by adapting conventional file system design to reduce Write Amplification. Although the related optimizations certainly improved performance, the authors of HiNFS suggest that there is still a performance bottleneck: poor write latency due to double-copy overheads in the kernel.
To resolve this, HiNFS proposes two optimizations: extensive \emph{latency hiding} behind the critical path and elimination of double copy overhead~\cite{ou-hinfs-2016}.\par
HiNFS achieves latency hiding by an \emph{NVMM-aware Write Buffer} policy. Using this policy, file writes are classified as \emph{eager}-persistent or \emph{lasy}-persistent. In the former, the write operation is performed immediately, without latency hiding. In the latter case, HiNFS buffers the request by moving its payload into a fast DRAM buffer. This $4$ kB-sized DRAM buffer uses the Least Recently Written (LRW) policy to order writes in temporal order. DRAM blocks are indexed by building a B-tree per file. A per-core kernel thread writes data from the DRAM buffer into PM, after which the DRAM blocks can be reclaimed. Write consistency is maintained by reusing PMFS's hybrid consistency mechanism.
\par
In the case of a read operation, HiNFS first checks if the corresponding blocks are in DRAM. When this is the case, the requested data can be returned immediately. If the data is not in DRAM, it is fetched from PM and directly copied into the user buffer. This avoids a double copy: device $\rightarrow$ kernel $\rightarrow$ user. Although this seems like a simple mechanism, it is more subtle. Suppose that a block is partly in DRAM, which may happen as PM is byte-addressable. In this case, HiNFS consults the \emph{cacheline bitmap}, which tracks the state of each $64$ byte cache line in a DRAM block, to determine which areas need to be fetched from PM.\par
The way HiNFS avoids double-copy overheads for write operations is quite complex; therefore, we provide only a brief explanation. Essentially, HiNFS uses multiple criteria (e.g., DRAM and PM latencies) to decide whether coalescing writes into one large write operation is beneficial. Periodically, these coalesced writes are moved to persistent memory.

\paragraph{Ext4-DAX} 
As we have seen in previous paragraphs, quite a bit of effort has gone into pushing PM support into the existing I/O stack. PMFS and HiNFS allow applications DAX to PM via a \texttt{mmap} interface. Instead of making extensive changes to the I/O stack, \emph{ext4-DAX} only adds DAX support to the existing \emph{ext4} file system~\cite{linux-foundation-direct-nodate}. We have already discussed DAX in~\autoref{sec:background-sw}. In short, applications can access PM without the interference of the kernel via CPU load and store instructions, bypassing the expensive OS page cache and generic block layer. In terms of throughput and latency, ext4-DAX performs similarly to PMFS~\cite{dulloor-pmfs-2014} and worse than HiNFS~\cite{ou-hinfs-2016}.

\par
In this section, we have seen four PM file systems that (extensively) modify the I/O stack to support PM. All four file systems are POSIX-compliant (see~\autoref{tab:fs-designs}), so applications can access PM without rigorous code changes. PBFS and PMFS introduced performant techniques to reduce Write Amplification: \emph{Short-Circuit Shadow Paging}, and atomic metadata updates. Additionally, they bypass the expensive OS page cache using DAX to PM. HiNFS addresses the double-copy overhead in the kernel to improve access latency. Ext4-DAX adds DAX support to the existing \emph{ext4} file system.

\subsection{Design: Contiguous File Allocation}
As mentioned before, file systems for block devices (e.g., \emph{ext4}) support large files using indirect blocks. An alternative is to position the file system and some of its data structures in virtual memory and take advantage of the hardware capabilities, the Memory Management Unit, better known as the \gls{mmu}. This approach has three advantages. First, keeping track of where a file is located is reduced to an operation involving two numbers: the starting address of the VMA and the offset within the file. Second, the number of seeks is minimized since the entire file can be read in one operation, instead of separate blocks~\cite{wu-scmfs-2013}. Third, DAX gives the user much more flexibility to design custom storage structures.

\par
Now, we will discuss the relevant designs proposed in the literature. We start by discussing SCMFS, a file system that inspires multiple modern PM file systems. Subsequently, we discuss file system designs that solve the limitations of SCMFS or propose other novel techniques, as seen in~\autoref{tab:fs-designs}.
\par
\begin{figure}[h]
    \centering
    \includegraphics[width=1.0\linewidth]{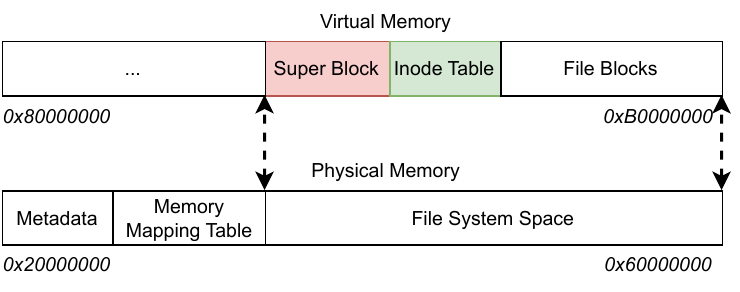}
    \caption{SCMFS physical and virtual memory layout}
    \label{fig:scmfs-layout}
\end{figure}

\paragraph{SCMFS}
To the best of our knowledge, the SCMFS file system~\cite{wu-scmfs-2013} was the first to implement a contiguous file system for Storage Class Memory (\gls{scm}), nowadays better known as PM. It uses the \gls{mmu} to map file addresses to physical addresses. \autoref{fig:scmfs-layout} displays a high-level view of SCMFS.
\par
The physical space contains the actual file system, its metadata, and a mapping table. This mapping table is used to initialize the MMU by inserting entries that map user space virtual addresses to physical addresses representing locations on the PM device. Modifications to these data structures are always transferred back to the PM device for consistency. Please note that both virtual and physical addresses need to be relative rather than absolute to account for randomization techniques such as \emph{Address Space Layout Randomization} (\gls{aslr})~\cite{marco-gisbert-address-2019}, which randomly arranges data in the virtual address space for security purposes.

\par
As depicted in~\autoref{fig:scmfs-layout}, the virtual space consists of three parts: the super block, the inode table, and the actual files. The super block serves the same purpose as in traditional file systems: keeping track of block/inode counts, block size, etc. ~\cite{linux-foundation-ext4-nodate}. The inode table stores file/directory metadata.
\par
Recall that the translation from virtual to physical addresses is done in hardware by the MMU. To speed up translation, the MMU includes a cache, the \emph{Translation Lookaside Buffer} (\gls{tlb})~\cite{li-ctfs-2022}. The TLB has a fixed number of address translation entries, usually between $16$ and $512$.
As this cache is quite small, we should use it efficiently to avoid the so-called \emph{TLB misses}.
Let us demonstrate this with an example. Consider a $2 MiB$ file mapped in virtual memory by $512$ conventional $4 kB$ pages, which may occupy all TLB entries~\footnote{For simplicity, we ignore that modern operating systems perform extensive software optimizations avoid poor TLB utilization~\cite{li-ctfs-2022, dulloor-pmfs-2014}}. Now suppose that we want to map multiple $2 MiB$ files in the file system space. In this situation, the number of cache misses increases dramatically, as both files cannot be in the TLB at the same time, increasing the latencies.
\par

SCMFS reduces potential TLB misses for large files by employing \emph{huge pages}. Instead of mapping one $2 MiB$ file using conventional $4 kB$ pages, it maps one $2 MiB$ huge page. This approach also avoids internal fragmentation; the amount of allocated but unused space.\par

Although SCMFS has shown good speed-ups in both (random) read and write workloads, it still has several limitations:

\begin{inparaenum}[(1)]
    \item In SCMFS, non-temporal data persistence and consistency are enforced by the \texttt{clflush} and \texttt{mfence} instructions, enforcing PM writes are temporally ordered. However, this is done at the expense of exposing the suboptimal write latency of PM devices to the critical path~\cite{ou-hinfs-2016, bhandari-implications-2012}.\label{pnum:2} 
    \item SCMFS does not avoid the costs of trapping into the kernel frequently: switching CPU protection modes, saving/restoring the trap frame, invoking the scheduler, etc.~\cite{sauthoff-costs-2021, baldini-soares-flexsc-2010}\label{pnum:3}
    \item Li et. al~\cite{li-ctfs-2022} note that SCMFS does not address the challenge of slow file resizing (using appends) and external fragmentation.\label{pnum:4}
\end{inparaenum}

We will address the issues related to consistency in~\autoref{sec:persistency-atomicity}. The other limitations are addressed by other PM file systems, which we discuss in a moment.

\par
\paragraph{SplitFS}
SplitFS~\cite{kadekodi-splitfs-2019} resolves two limitations of SCMFS, namely kernel trap overhead and append performance. To accomplish this, SplitFS proposes a \emph{hybrid} design: a file system design in which user and kernel space have distinct responsibilities. A user space library (\emph{U-Split}) services the \emph{data path}, that is, all data operations in virtual memory, e.g.: \texttt{read()}, \texttt{write()}, or an append. Metadata-related operations, for example, \texttt{open()} or \texttt{rename()}, are handled by a kernel library (\emph{K-Split}), i.e. the \emph{control path}.

\par
\begin{figure}[h]
    \centering
    \includegraphics[width=1.0\linewidth]{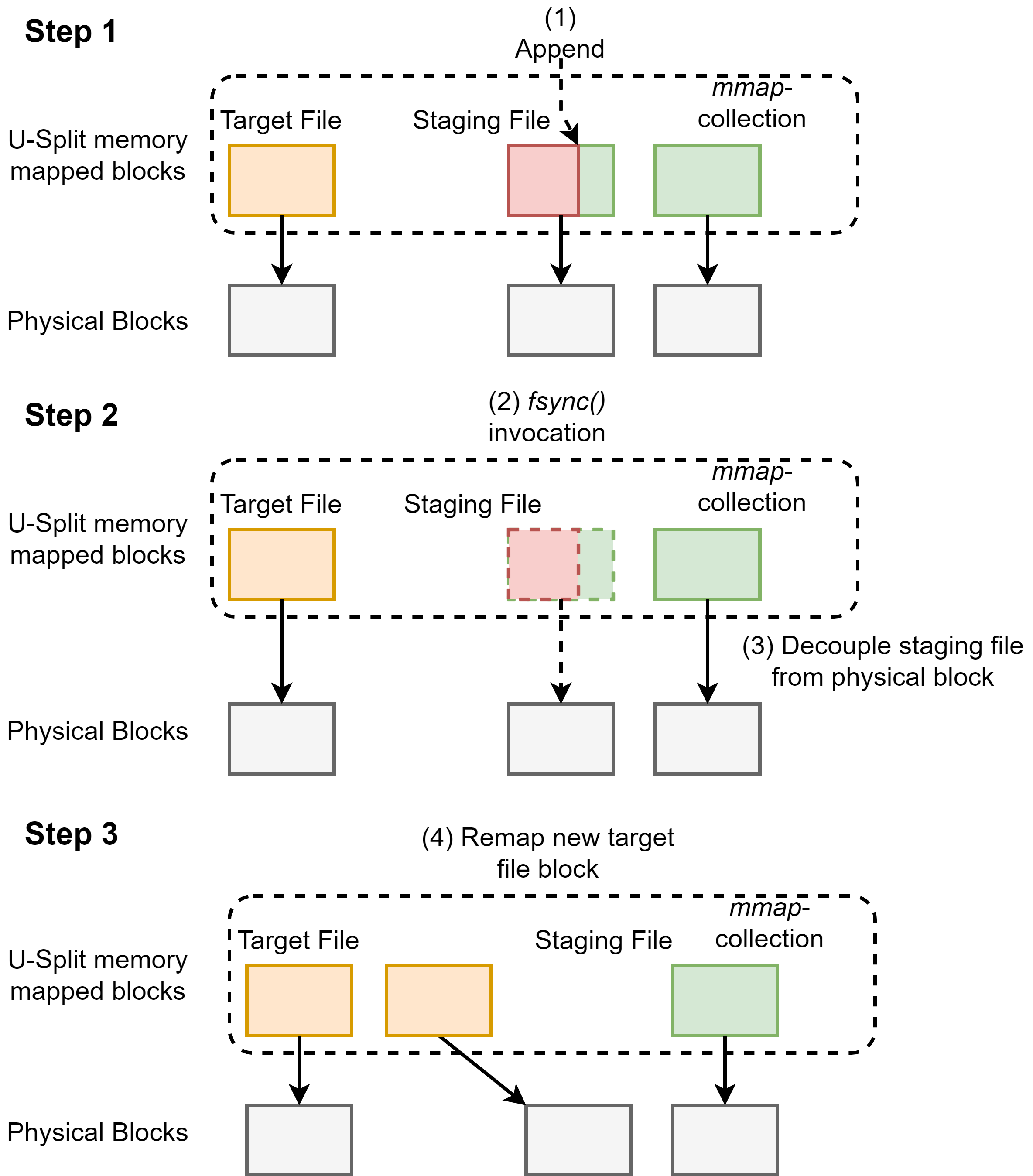}
    \caption{SplitFS U-Split \emph{relink} procedure, partly based on the figure made by Kadekodi et al.~\cite{kadekodi-splitfs-2019}}
    \label{fig:usplit-relink}
\end{figure}

The main performance-degrading aspect of an append operation is data copying in the kernel~\cite{kadekodi-splitfs-2019}. To mitigate this issue, the authors of SplitFS propose the use of \emph{staging files}. Instead of appending to the actual file, the operation is redirected to a temporary staging file in PM, managed by U-Split; see~\autoref{fig:usplit-relink}. Note that in this case file data may be spread over two locations, the stage and actual files. Consequently, U-Split maintains a collection of memory-mapped areas per file for accounting purposes.
\par
Eventually, all staged file modifications must be made persistent. In U-Split, this flushing procedure is initialized after capturing a POSIX \texttt{fsync()} using \texttt{LD\-PRELOAD}. Then, a \emph{relink} procedure is executed. This procedure logically moves the PM blocks from the staging file to the target file by a \emph{zero-copy operation}: an operation that avoids unnecessary data copies~\cite{tianhua-design-2008}. The relink procedure involves several steps, as shown in~\autoref{fig:usplit-relink}: 

\begin{inparaenum}[(1)]
    \item In the first step, an append operation is performed. This operation is redirected to a staging file. Note that each staging file is mapped to a physical block, and the auxiliary translation is performed by the MMU, as in SCMFS.\par
    \item Eventually, \texttt{fsync()} is invoked by the user, so the flushing procedure starts. One block of the staging file is decoupled from its corresponding physical block.\par
    \item \& \item This physical block is relinked with the target file to form a new contiguous file in the VMA.
\end{inparaenum}

Note that in the aforementioned steps, data copies are avoided. We modify only the virtual-to-physical page mapping in the page table, which is a relatively inexpensive operation.\par

In terms of performance, SplitFS achieves good throughput: $27\%$ improvement in sequential reads and $7.85\times$ speedup in appends. Metadata-heavy workloads still introduce an overhead of up to $13\%$ due to the additional bookkeeping required to manage the staging files.

\paragraph{Aerie}
Aerie is another file system that implements the hybrid user/kernel space architecture. Compared to SplitFS, Aerie shifts even more responsibility to user space, improving metadata-heavy workloads. Applications can define their own workload-specific file system interfaces in user space. This design allows for higher performance than a single generic interface, e.g., POSIX~\cite{kadekodi-splitfs-2019}. Aerie exposes two services in user space: \emph{libFS} and the \emph{Trusted File System Service}, and the \emph{SCM Manager} in kernel space. 
\par
LibFS provides applications with the essential ability to define a file system: the ability to map file names to file metadata, and indexing, which translates a file offset into a byte in memory. The Trusted File System Service (TFS) handles the integrity of metadata updates and concurrency. It runs as a Remote Process Call (\gls{rpc}) service accessible through a user-mode process. The SCM Manager multiplexes physical PM allocations and maps backing pages into user space. Note that for compatibility purposes, Aerie also defines a POSIX-like interface, PXFS.
\par
The performance of Aerie is evaluated using PXFS. Compared to kernel-mode file systems (i.e., ext4), PXFS achieves $53 \% \sim 109 \%$ higher throughput in single-threaded workloads. Unfortunately, the freedom Aerie provides comes at the cost of poor multi-core scalability. PXFS do not scale linearly beyond four threads due to contention in the TFS's storage allocator.

\paragraph{Kuco}
The authors of Kuco~\cite{dong-performance-2019} find that multicore scalability has not been well addressed by other file systems. For example, Aerie relies on a centralized TFS to enforce concurrency control, which becomes a bottleneck in high-concurrency workloads. File systems that avoid the page cache, e.g. SCMFS and SplitFS, still experience software overhead due to kernel traps and the VFS.
\par
Kuco's high-level design is very similar to that of SplitFS, in the sense that it proposes a \emph{client/server} model in which user (\emph{Ulib}) and kernel space (\emph{Kfs}) have their own responsibilities. Kuco's design shifts even more responsibility to the user space to decrease the involvement of the kernel.
To do this, Kuco introduces three new techniques: \emph{collaborative indexing}, \emph{two-level locking}, and \emph{versioned read}.
\par

\emph{Collaborative indexing}: Kuco offloads most of the path name resolution to a user space library, \emph{Ulib}. Applications communicate with a dedicated \emph{Ulib} instance to perform metadata operations, e.g., \texttt{chown()}. An Ulib instance looks up file metadata through the so-called \emph{partition trees}. These data structures are quite complex; therefore, we will not elaborate on all the details here. For now, it is sufficient to understand that these complex tree structures maintain all the file/directory inodes contained in a file system partition. 
When a metadata operation is initiated, \emph{Ulib} performs the required pathname resolution by traversing the corresponding partition tree. It looks up all related metadata items in user space and includes the associated virtual memory pointers in the system call payload.  After performing the necessary consistency checks, \emph{Kfs} can perform the metadata operations directly by writing to the corresponding virtual memory. Using this approach, expensive locking is avoided as only \emph{Kfs} can perform metadata updates.
\par
\emph{Two-level locking} is used to coordinate concurrent file writes. Kuco introduces \emph{direct access range-lock} to serialize fine-grained concurrent writes by performing region locking. Using this lock, multiple threads can write different data pages in the same file simultaneously. 
\par
The \emph{versioned read} mechanism allows for user-level reads without any kernel involvement, avoiding an expensive RPC or system call. It ensures that readers never read data that is out-of-date/incomplete by embedding a `version field' in each data pointer inside the block mapping table.
\par
Combining these techniques, Kuco achieves up to one magnitude higher throughput in small I/O workloads compared to SplitFS, PMFS, and Ext4-DAX. In a $16$-thread Filebench~\cite{mcdougall-filebench-2004} benchmark, Kuco outperforms PMFS throughput by $1.2\times$, and Ext4-DAX by $1.9\times$.

\paragraph{ZoFS}
The last hybrid file system that we discuss is ZoFS. Like Aerie, ZoFS gives the user space direct control over both data and metadata, allowing applications to design their own file systems. We already mentioned that Aerie's multicore performance does not scale linearly due to contention inside the user space TFS. To improve parallel performance, ZoFS proposes an alternative implementation in which applications can access metadata without a user-space library.
\par
In Aerie, applications must request permission from the TFS every time it wants to access file data. In ZoFS, an application only requests permission once, avoiding TFS's performance-degrading Remote Procedure Calls (\gls{rpc}s). This is done by issuing a system call directed to ZoFS's kernel module: \emph{KernFS}. If permission is granted,  KernFS assigns the application an \emph{coffer}: a range of PM pages that share the same permission properties~\cite{dong-performance-2019}. Protection and isolation of a \emph{coffer} are enforced in hardware through Intel's Memory Protection Keys (MPK)~\cite{corbet-memory-2015, noauthor-intel-2022}. It allows the kernel to restrict the permissions of memory regions mapped in user space. For example, one could map a set of PM pages as read-only in user space. This feature is supplemental to the MMU's page protection bits; both permission checks will be performed during memory access. 
Once access is granted and associated \emph{coffer} PM mappings are inserted, applications can access the associated memory region without any kernel interference during the application lifetime. 
\par
Compared to other PM file systems, ZoFS achieves higher throughput in workloads that affect a fixed set of file blocks, e.g., constantly appending data to the same set of files. This is because the number of context switches into the kernel is significantly reduced. Workloads that involve a dynamic set of files perform worse due to an increase in \emph{coffer} access requests, resulting in a higher number of context switches.
\par
To summarize, a contiguous file system design enables applications to access PM from user space, reducing kernel involvement. These `hybrid' file systems (SplitFS, Aerie, Kuco, and ZoFS) allow the construction of application-tailored file systems in user space, reducing the role of the kernel, which ultimately results in better parallel performance and throughput compared to PM file systems that extensively modify the existing kernel I/O stack (see \autoref{sec:trad-design}).

\subsection{Design: Log-Structured}
This category of PM file systems mainly uses logs for storage. As mentioned in~\autoref{sec:crash-consistency-techq}, log-structured file systems optimize for sequential performance and are still relevant today. Modern log-structured file systems for flash devices are SFS~\cite{min-sfs-2012} and F2FS~\cite{lee-f2fs-2015}. SFS proposes the \emph{cost-hotness policy}, where blocks with similar 'hotness' are assigned to groups for faster Garbage Collection. F2FS introduces multi-head logging: data blocks are categorized as "cold", "warm", or "hot" and are located at separate physical \emph{zones} on the flash device.
SFS and F2FS are designed to work with a wide range of flash devices, e.g. conventional block SSDs or Zoned Namespace (\gls{zns}) devices. Therefore, they do not take advantage of the byte-addressable properties of PM.
\par
In this subsection, we discuss NOVA~\cite{xu-nova-2016} and Strata~\cite{kwon-strata-2017}, log-structured file systems that take advantage of PM to create faster log-structured file systems.

\paragraph{NOVA}

The log-structured NOVA file system~\cite{xu-nova-2016} aims to maximize PM performance while providing stronger consistency guarantees than BPFS~\cite{condit-better-2009}, SCMFS~\cite{wu-scmfs-2013}, and Aerie~\cite{volos-aerie-2014}. As mentioned before, BPFS does not perform well in certain operations, e.g. a directory move. Furthermore, SCMFS does not provide any consistency guarantees for (meta) data. Another observation they make is that Aerie does not support atomic data operations. 
\par
Conventional log-structured file systems store metadata and actual data blocks in the log. NOVA deviates from this design in three design choices.
\par
First, it proposes a logging structure where each inode is assigned a separate log to allow for parallelism. Synchronization primitives, e.g., mutexes or locks, are avoided by only allowing one open transaction at a time on each core. Logs are stored as linked lists in PM, so they can grow or shrink in length and do not need to be contiguous in memory. 
\par
Second, logs only store file metadata, so no data blocks. The data blocks are divided into pools, one per CPU. Each CPU maintains a red-black tree~\cite{morris-data-nodate} in DRAM to keep track of free blocks in ascending order of addresses to enable fast merging and deallocation.
\par
Third, NOVA uses journaling in cases where metadata updates span multiple inodes. For example, changing file permissions is a metadata-only operation; however, truncating a file updates the file data, file size, and modification date. In such cases, NOVA first commits the data. Then, the updated metadata is appended to the corresponding file inode log. Finally, NOVA journals all affected log tails.
\par

\begin{figure}[h]
    \centering
    \includegraphics[width=1.0\linewidth]{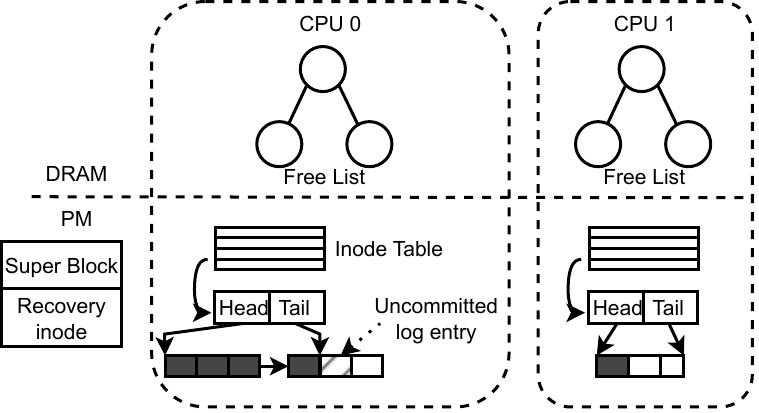}
    \caption{NOVA file system log-structure}
    \label{fig:nova-log-structure}
\end{figure}

\autoref{fig:nova-log-structure} graphically displays the layout of the NOVA file system. As mentioned before, note that each CPU maintains its own free list and inode table. Each table entry stores two pointers; one for the inode log head and tail, respectively. Additionally, NOVA stores a super block and a recovery inode. The first has a purpose similar to that of conventional file systems. The latter stores the page allocator state to allow faster recovery after normal shutdown.
\par
In case of an improper shutdown, NOVA performs a recovery routine, consisting of two steps. First, NOVA rolls back any uncommitted file system transactions to bring the file system back into a consistent state. Second, each CPU reconstructs its free list by scanning the inode table, also known as a \emph{log scan}.

\paragraph{Strata}
Another log-structured file system is Strata. The file system model of Strata is fundamentally different compared to the file systems we discussed before. These file systems assume that each file system is linked to a single physical device, in this case, PM. Strata, on the other hand, proposes a file system that spreads data across different storage devices. This enables Strata to capture the unique properties of multiple storage devices, such as PM, SSD, or HDD, in one file system~\cite{kwon-strata-2017}.
\par
\begin{figure}[h]
    \centering
    \includegraphics[width=0.9\linewidth]{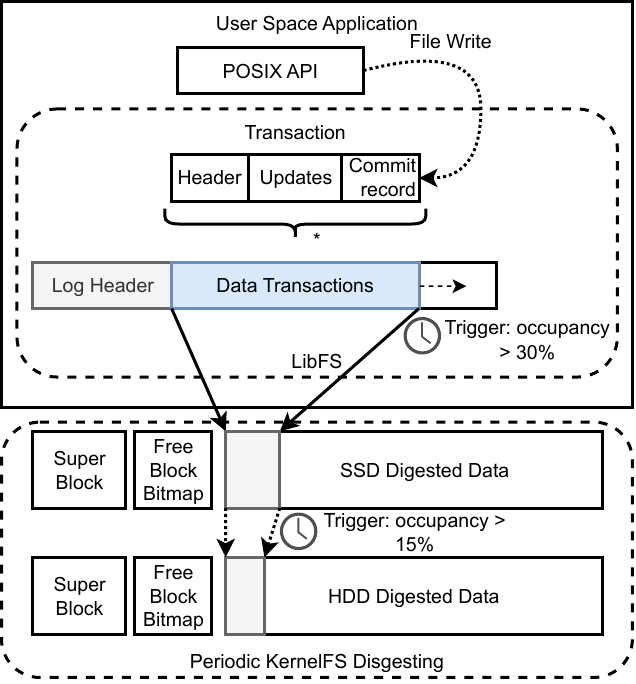}
    \caption{Strata file system log-structure and data digesting, triggered by log occupancy}
    \label{fig:strata-log-structure}
\end{figure}

To implement such a storage model, Strata implements a split architecture similar to what we have seen in SplitFS~\cite{kadekodi-splitfs-2019}, as displayed in~\autoref{fig:strata-log-structure}. We will elaborate the most important components in this figure.

\emph{LibFS} exposes a POSIX-compliant interface to applications in user space. To attach fast write performance, each application is assigned a dedicated log for file system I/O operations, stored in fast PM. Note that this multi-log design differs from what we have seen in the NOVA FS; each log is assigned a separate log, instead of one log per file.\par
Applications access Strata through POSIX calls. File writes are transformed into transactions and appended to the log, as depicted in~\autoref{fig:strata-log-structure}. File reads are handled with the help of an inode-like data structure, which we will discuss in a moment.
\par
In the kernel space, \emph{KernelFS} is responsible for garbage collection and \emph{digesting}: the process of aggregating file data into sequential disk areas to minimize fragmentation~\cite{kwon-strata-2017}. This means that in user space, where writes are not block-aligned, digesting ensures that device-level write amplification is minimized by coalescing writes into block-aligned writes favorable for the selected 'level'. KernelFS maintains multiple digest levels, where each level corresponds to a single storage device, as illustrated in~\autoref{fig:strata-log-structure}. Digesting is performed in the background and is initiated when the application log is filled beyond a threshold, for example, $30\%$.
\par
Due to digesting, the file's data blocks may be scatted over multiple digest levels, complicating file reads. Therefore, Strata uses adapted inodes to structure file metadata. Each inode contains one or more \emph{extent trees}, each representing a storage device. The tree nodes point directly to the file's data blocks. \par

\paragraph{Discussion}

\begin{figure}[h]
    \centering
    \includegraphics[width=1.0\linewidth]{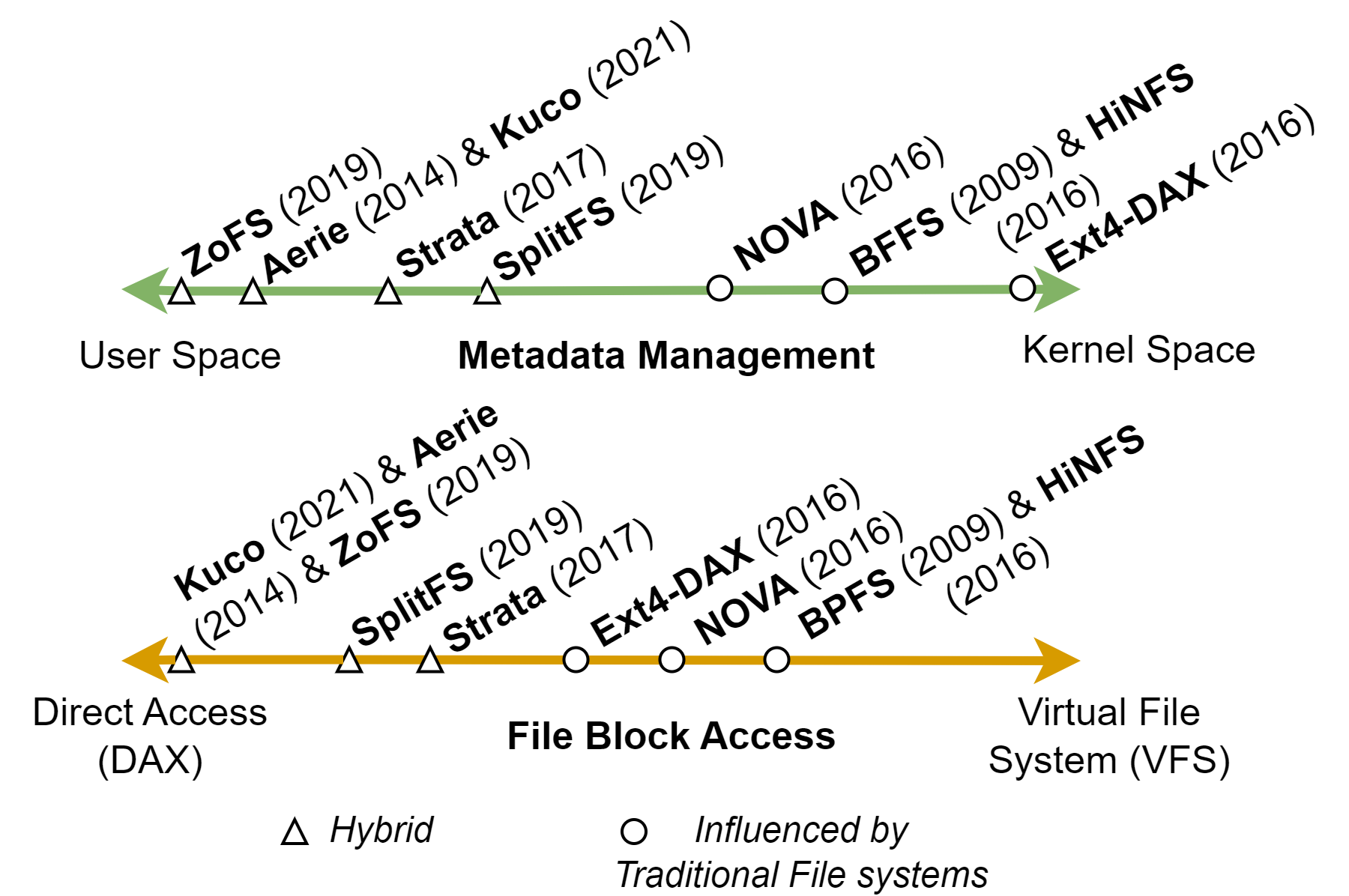}
    \caption{PM file systems positioning: \emph{Metadata Management} and \emph{File Block Access}}
    \label{fig:pm-filesystems-timeline}
\end{figure}

In~\autoref{sec:trad-design}, we discussed the first PM-capable file systems released between 2009 and 2016 (\autoref{fig:pm-filesystems-timeline}). These file systems adapt the existing kernel I/O infrastructure to support PM. Additionally, they use DAX to bypass the OS page cache. Hybrid file systems use a different approach. These file systems enable the construction of PM-aware file systems in user space, allowing application-tailored optimizations. In addition, they reduce kernel overhead by offloading metadata management to user space, increasing throughput and parallel performance. 
This class of PM-file systems became mainstream, as seen in~\autoref{fig:pm-filesystems-timeline}.

\section{File Indexing Overhead} \label{sec:file-indexing-overhead}

As discussed in the introduction (\autoref{sec:introduction}), PM achieves performance close to DRAM, shifting the overhead from the device to the I/O stack. Multiple studies show that file indexing has a significant impact on performance~\cite{li-ctfs-2022, neal-hashfs-2021, wang-byvfs-2018}; in extreme cases, up to $45\%$ of the total runtime~\cite{cai-flatfs-2022}. In this section, we consider the two most prominent issues related to file indexing when using PM: \emph{file mapping} and the \emph{path walk} overhead in the Linux Virtual File System (VFS).
\par
HashFS and ctFS aim to improve \emph{file mapping} performance, that is, the operation of mapping a logical file offset to a physical location on the underlying storage device~\cite{neal-hashfs-2021}. HashFS uses a \emph{hash table} to extract more performance, while ctFS proposes a design in which the translation is performed in hardware using the MMU.
\par
Afterward, we cover FlatFS and ByVFS, which decrease VFS path walk overhead. In a path walk, the VFS traverses a file path, for example, \texttt{"/foo/bar.txt"}, to return information about the file \texttt{bar.txt}.

\subsection{Improving File Mapping Performance}

The authors of HashFS~\cite{neal-hashfs-2021} mention that multiple properties negatively impact the performance of a PM file system. First of all, file system fragmentation results in files being allocated in non-contiguous PM regions, resulting in larger mapping structures, which in turn causes poor search and insert performance. Take PMFS's B+ tree (\autoref{sec:file-system-design}), for example. If a file is heavily fragmented the tree grows rapidly in size, resulting in slower tree traversal~\footnote{A B+ tree worst-case time complexity is logarithmic: $O(log \text{ } n)$~\cite{braginsky-lock-free-2012}}.\par
Another issue is related to the per-file mapping scheme. Many traditional file systems optimized for slow block devices (e.g.~\emph{ext4}) use a simple isolation mechanism: a read/write lock covers the entire structure to enforce mutual exclusion. Consequently, concurrent performance is limited; only one thread may write to a file in any given time frame. 
\par
HashFS' main contribution is to solve the aforementioned issues by implementing hash-based file mapping.  Compared to Radix/Extent trees (see~\autoref{sec:data-structures}), hash tables require considerably fewer and smaller memory accesses, therefore, are convenient to store in PM. It is important to understand that HashFS should be seen as a standalone optimization; its contributions can be used in unison with existing state-of-the-art PM file systems that we discussed in~\autoref{sec:file-system-design}.
\par

The HashFS paper describes two implementations. The first implementation uses \emph{Cuckoo} hashing; a form of hashing in which each entry is hashed twice using two different hash functions to avoid hash collisions~\cite{pagh-cuckoo-2004}. \autoref{fig:cuckoo-hash-table} displays an example in which a Cuckoo hash table lookup is performed. It consists of multiple steps:

\begin{figure}[h]
    \centering
    \includegraphics[width=1.0\linewidth]{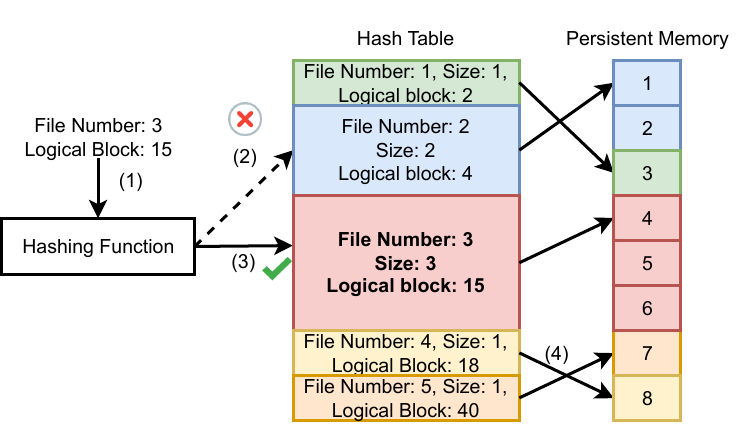}
    \caption{Cuckoo hash table lookup example}
    \label{fig:cuckoo-hash-table}
\end{figure}

\begin{inparaenum}[(1)]
    \item In the first step, two different hashing functions compute the hash for the logical block number $15$.
    \item The first hash points to the second entry in the hash table; however, its logical block is not equal to $15$.
    \item The second hash points to the correct entry as its logical block matches.
    \item We find the corresponding physical block number by consulting the metadata structure stored in the hash table entry.
\end{inparaenum}
\par

The second (final) HashFS implementation uses \emph{linear probing}: a hashing function that avoids hash collisions by maintaining key-value pairs for each hash table entry that contains conflicting hash values. Compared to Cuckoo hashing, linear probing limits search overhead in the event of a hash collision, as conflicting entries are stored in adjacent locations, which is beneficial for PM performance~\cite{swanson-early-2019}. The operation of mapping a logical file offset to a physical PM location consists of multiple steps, as depicted in~\autoref{fig:hashfs}:

\begin{figure}[h]
    \centering
    \includegraphics[width=1.0\linewidth]{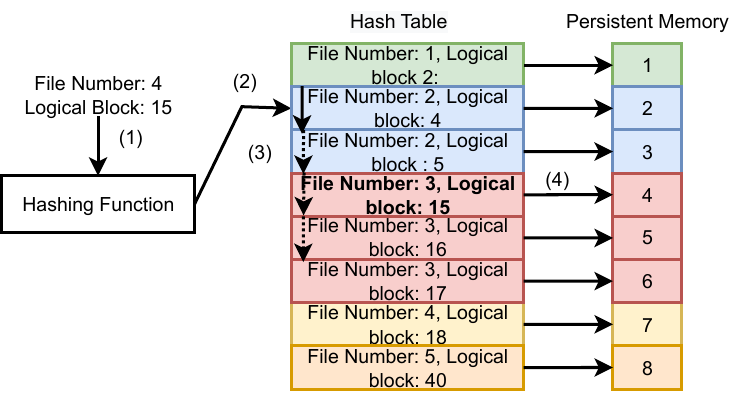}
    \caption{HashFS file mapping operation using \emph{Linear Probing}}
    \label{fig:hashfs}
\end{figure}

\begin{inparaenum}[(1)]
    \item First, the logical block number $15$ is hashed.
    \item In the second step, we jump to the corresponding entry in the hash table. The logical block does not match, so a hash collision must have occurred.
    \item We iterate over the conflicting entries in adjacent locations until we have found a match.
    \item As the hash table entries and the physical PM locations are mapped one-to-one, the entry offset in the hash table is the physical block number.
\end{inparaenum}

This one-to-one mapping scheme makes it straightforward to achieve good parallel performance: approximately a $4.5 \times$ decrease in latency for $4$ kB sequential reads, $5 \times$ decrease for $4$ kB random reads, and $10 \times$ decrease for $4$ kB inserts compared to Radix Trees.
\par

ctFS takes a different approach by offloading file mapping operations to the MMU~\cite{li-ctfs-2022}. In a Background section (\autoref{sec:background-sw}), we already mentioned the notion of a \emph{hierachical page table}: a tree structure that maps virtual addresses to physical addresses.  Note that a hierarchical page table and a Radix tree (see~\autoref{sec:data-structures}) share a common feature; they both use simple offset calculations to traverse the tree. A key difference is that the Radix tree walk is implemented in software, whereas the faster page table walk is implemented entirely in hardware. Therefore, offloading the file mapping operation to the MMU could be very beneficial for performance.\par

ctFS introduces the notion of a \emph{Persistent Page Table} (\gls{ppt}): a page table that can be stored in PM. Conventional page tables are volatile structures, which means that, at shutdown, they are lost forever. In the context of performing a file mapping operation, this is undesirable behaviour, as logical to physical block mappings would be irreversibly lost after shutdown. Therefore, ctFS stores the page table entries used for file mapping entirely in PM. 
During initialization, ctFS copies PTT entries into the kernel's DRAM page table, which is then used by the MMU to perform fast address translation when accessing PM. In the event of a page fault, ctFS allocates a new persistent page, creates a new mapping in the PTT, and finally copies the mapping to the kernel page table.
\par
Like SplitFS, ctFS uses a hybrid architecture where user and kernel space have distinct responsibilities. \emph{ctU} manages the structure of the file system. It maintains various \emph{partitions}. Each partition level contains blocks that are $8\times$ the size of the partitions in the previous level, for example, $32$ kB in level 1, since the partition size of level 0 is $8$ kB.
\emph{ctK} makes sure that the MMU can perform the address translation and that the changes are persistent. It does so by ensuring that the kernel' DRAM page table mappings are an exact copy of those stored in the PPT in PM.
\par
\begin{figure}[h]
    \centering
    \includegraphics[width=1.0\linewidth]{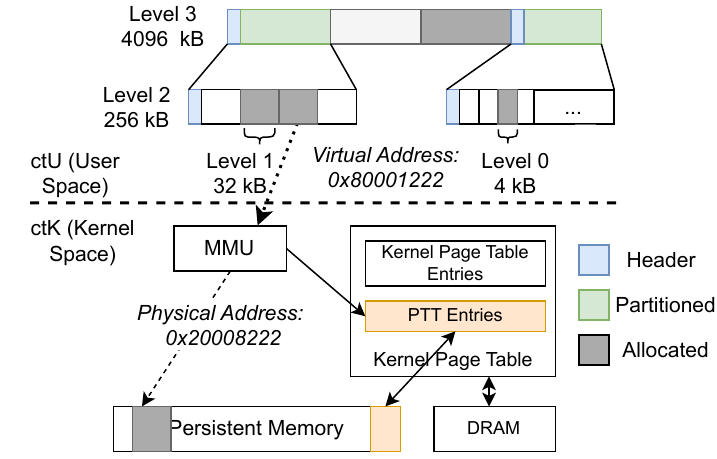}
    \caption{ctFS file mapping operation example}
    \label{fig:ctfs-structure}
\end{figure}

\autoref{fig:ctfs-structure} provides an example in which a file mapping operation is performed. Observe that the entire file system is mapped into user space. Suppose a user performs a read operation at virtual address \texttt{0x80001222}. To find the corresponding physical location in PM, the MMU performs a page table walk. More specifically, the MMU refers to the kernel page table to find the corresponding PTT entry, which contains the virtual-to-physical memory mapping, in this case \texttt{0x80001222} to \texttt{0x20008222}.

In short, we have seen two techniques to improve file mapping performance. Hash tables are convenient structures to store in PM and allow for fine-grained locking, resulting in improved parallel performance. Offloading the translation to the MMU is another promising technique. 

\subsection{Virtual File System Overhead}

The Virtual File System (VFS) provides a software abstraction that enables access to different file systems using a uniform interface. In addition, it provides protection and concurrency.\par
The VFS caches multiple structures to improve performance, namely: the super block, directory entries (\emph{dentry}), and file inode. The latter two are especially important for the \emph{path walk}. During the walk, the VFS inserts the corresponding intermediate directory entries into a \emph{dentry} cache~\cite{kernel-development-community-pathname-nodate}. This is beneficial for performance as applications are likely to access the same file again in the near future~\cite{wang-byvfs-2018, cai-flatfs-2022}.\par
In traditional slow block devices, e.g. an HDD, disk latencies are significantly higher than path walk latency. In such cases, caching intermediate directory entries is the logical thing to do. However, this claim does not apply to PM, as the time spent retrieving the entry from the \emph{dentry} cache is greater than the latencies observed when writing/reading to PM, decreasing overall performance.
\par
Based on this insight, the authors of ByVFS~\cite{wang-byvfs-2018} suggest that removing the \emph{dentry} cache would be beneficial for performance, especially in small I/O workloads. They claim to reduce the execution time by $\sim\ 48\%$ for a \emph{stat} call, using NOVA as the underlying file system.
\par
The authors of the FlatFS paper~\cite{cai-flatfs-2022} identified another issue. They acknowledge that ByVFS improves indexing performance significantly; however, they point out another issue related to the path walk. The files and directories contained in the namespace tree are physically scattered throughout the storage device, resulting in poor data locality and indirect memory accesses~\cite{cai-flatfs-2022}. Additionally, the namespace tree traversal introduces random memory accesses as the directory entries of different directories are scattered across the device. Multiple studies~\cite{gugnani-understanding-2021, yang-empirical-nodate} have shown that such random access patterns result in suboptimal PM performance. 
\par
To solve these problems, FlatFS proposes the novel \emph{coordinated file path model}. This model improves data locality performance by introducing a `flat' namespace structure in which contiguous directory entries in a namespace are also stored consecutively within PM. A path walk only involves a single lookup, avoiding the aforementioned expensive namespace tree traversal.
\par

The `coordinated file path model' consists of two components: the traditional component-at-a-time model and the novel full-path-at-a-time model. The traditional walk model is included to accommodate namespace switches. Such a namespace switch may occur when certain semantic path components (e.g., dot-dot ($..$), a mount point change, or symbolic links) are encountered.\par
The `full-path-a-at-a-time model' first performs preprocessing to speed up the lookup, e.g., dots, redundant slashes, or semantic path elements are removed. After preprocessing, this canonical path is passed to the \emph{semantic path component finder}. This component searches for the corresponding inode in a persistent range-optimized $B^r$ tree, after which a permission check occurs. A $B^r$ tree is an adapted B tree that provides faster range operations~\cite{cai-flatfs-2022}, speeding up file system operations, such as a directory copy. Its tree nodes are $256$ bytes aligned, the optimal memory access granularity of Intel Optane DC Persistent Memory~\cite{yang-empirical-nodate}.
A $B^r$ tree lookup is performed using the \emph{Write-optimized Compressed} (\gls{woc}) key as an index. Instead of using the full preprocessed path as an indexing key, FlatFS uses the smaller WoC key to reduce storage consumption. Every index key is divided into two parts, a prefix and a suffix. All keys in the same tree node share the same prefix, allowing for smaller index keys.
\par
In addition, WoC keys use a complex caching layout to avoid write amplification. Suppose that we insert a new key into the $B^r$ tree. This insertion may cause prefix expansion of other keys, leading to many small writes. Instead of storing the entire prefix within each key, FlatFS caches prefixes in DRAM and only adjusts the \emph{prefix} size when necessary, avoiding write amplification and costly cache line flushes.
\par
Using the aforementioned techniques, FlatFS achieves stable path latency, regardless of the file path length. Moreover, FlatFS outperforms a hot \emph{dentry} cache, which strengthens the claim that a directory cache imposes a performance penalty when using PM.

\par
To summarize, we have seen that a substantial amount of VFS overhead comes down to caching and an expensive path walk. Due to PM's low read latency, some caching structures inside the VFS serve no purpose, in particular the directory cache. Path walk performance can be improved by a coordinated file path walk model.

\section{Data Crash-Consistency} \label{sec:persistency-atomicity}

In this section, we investigate how persistent memory file systems guarantee data crash consistency. First, we classify the PM file systems, discussed in Sections~\ref{sec:file-system-design} and~\ref{sec:file-indexing-overhead}, by the high-level consistency technique(s) they employ, i.e., as shadow paging, log-structuring, journaling, or more exotic/hybrid variants. We already provided the relevant background for the conventional well-known persistence techniques in~\autoref{sec:crash-consistency-techq}. Therefore, we focus on more novel/innovative designs that provide persistence guarantees while maintaining good performance. 
Subsequently, we address the issue of \emph{write reordering}.
\par
This combination of data consistency and enforced write ordering allows PM file systems to support ACID transactions, that is, transactions that adhere to the properties of Atomicity, Consistency, Isolation, and Durability~\cite{gu-pisces-2019}. An operation is \emph{atomic} if and only if updates are committed in all or none manner~\cite{dulloor-pmfs-2014}. Data \emph{consistency} implies that memory writes must be performed in a strict format/order for correct recovery in the event of a crash. \emph{Isolation} ensures that transactions do not affect each other. \emph{Durability} ensures that the data affected by a completed transaction must be persistent, even in the event of system failure~\cite{lu-loose-ordering-2014, wright-extending-2007}.

\subsection{Consistency Techniques}

\autoref{tab:crash-consistency-techniques} provides an overview that relates different crash-consistency techniques to the PM file systems we discussed in previous sections. Based on this overview, we can derive multiple interesting trends.

\begin{table*}[h]
    \centering
    \small
    \begin{tabular}{|l|P{0.18\textwidth}|P{0.18\textwidth}|P{0.18\textwidth}|P{0.18\textwidth}|}
        \hline
      & Atomic in-place updates & Log-structuring & Shadow Paging & Journaling \\
        \hline
        BPFS & & & \cellcolor{green!30}$\checkmark$ \emph{Short-Circuit Shadow Paging} & \\
        \hline
        PMFS & \cellcolor{orange!30}$\checkmark$ Small Metadata Updates & & \cellcolor{orange!30}$\checkmark$ Only for Data Blocks & \cellcolor{orange!30}$\checkmark$ Undo Logging, for Large Metadata Updates \\
        \hline
        HiNFS & \cellcolor{orange!30}$\checkmark$ Small Metadata Updates & & \cellcolor{orange!30}$\checkmark$ Only for Data Blocks & \cellcolor{orange!30}$\checkmark$ Undo Logging, for Large Metadata Updates \\
        \hline
        Ext4-DAX & & & & \cellcolor{green!30}$\checkmark$ Redo Logging* \\
        \hline
        SCMFS & & & & \\
        \hline
        SplitFS & & & \cellcolor{orange!30}$\checkmark$ Using \texttt{relink} primitive & \cellcolor{orange!30}$\checkmark$ Operation Logging to record file operations \\
        \hline
        Aerie & & & & \cellcolor{green!30}$\checkmark$ Redo Logging \\
        \hline
        NOVA & \cellcolor{orange!30}$\checkmark$ Small Metadata Updates & \cellcolor{orange!30}$\checkmark$ Large Metadata Updates & & \cellcolor{orange!30}$\checkmark$ Metadata Updates spanning multiple inodes \\
        \hline
        Strata & & & & \cellcolor{green!30}$\checkmark$ Operation Logging in user space, Redo Log for digest areas in kernel space  \\
        \hline
        ctFS & & & \cellcolor{orange!30}$\checkmark$ Only for Data Blocks, using \texttt{pswap} primitive & \cellcolor{orange!30}$\checkmark$ Redo Logging for Metadata Updates  \\
        \hline
        Kuco & & & \cellcolor{orange!30}$\checkmark$ Only for Data Blocks & \cellcolor{orange!30}$\checkmark$ Operation Logging to record file operations \\
        \hline
        ZoFS & \cellcolor{green!30}$\checkmark$ Small Metadata Updates & & & \\
        \hline
        HashFS & \cellcolor{green!30}$\checkmark$ Hash Table Insertions & ** & ** & ** \\
        \hline
    \end{tabular}
    \caption{PM file systems crash-consistency techniques. Orange cells represent hybrid implementations. *: optional, **: HashFS only implements PM-optimized file mapping, file block consistency should be enforced through the PM file system~\cite{neal-hashfs-2021}.}
    \label{tab:crash-consistency-techniques}
\end{table*}

\begin{inparaenum}[(1)]
    \item First, note that several file systems (PMFS~\cite{dulloor-pmfs-2014}, HiNFS~\cite{ou-hinfs-2016}, NOVA~\cite{xu-nova-2016}, ctFS~\cite{li-ctfs-2022}, Kuco~\cite{chen-kuco-2021}, ZoFS~\cite{dong-performance-2019}) use a hybrid approach. In this approach, small (meta) data updates are performed using atomic in-place updates, whereas more performance-degrading techniques handle larger writes, for example, journaling~\cite{condit-better-2009}. Suppose that we want to append to a file. Instead of inserting a new log entry into the journal consisting of all modifications performed, e.g. data blocks, file modification date, file size, etc., we distinguish between small and large writes. Small writes, such as modifying the file modification date or size, can be performed using low-cost atomic instructions, e.g., Intel's 64-byte atomic store instructions~\footnote{See Section 8.2.4 - Intel Architecture Software Developer Manual~\cite{noauthor-intel-2022}.}. The atomicity and durability of larger writes is enforced through the more performance-diminishing crash-consistency techniques we discussed in~\autoref{sec:crash-consistency-techq}: \emph{Journaling}, \emph{Shadow Paging}, \emph{Log-Structuring}.
    \par
    
    \item Second, an emerging trend is that PM file systems use \emph{operation logging} (\autoref{sec:crash-consistency-techq}) to record transactions. Bhat et al.~\cite{bhat-scaling-2017}, authors of ScaleFS, have shown that an in-memory file system, in combination with an operation log, results in reduced write amplification and improved concurrent performance on slow disk devices compared to more conventional undo/redo logging. These forms of logging involve either a full copy to record data in the log (Ext4-DAX), perform a lazy copy using Shadow Paging (BPFS), or require extensive modification of file system metadata structures (PMFS, HiNFS)~\cite{kwon-strata-2017}.  SplitFS~\cite{kadekodi-splitfs-2019}, Strata~\cite{kwon-strata-2017} and Kuco~\cite{chen-kuco-2021} bring the concept of operation logging to PM file systems. 
    \par
    In SplitFS and Kuco, the log entries do not contain file data; instead, they only contain a pointer to the staging file in memory~\cite{kadekodi-splitfs-2019}. Log entries persist by performing 64-byte atomic writes. To ensure that log entries persist in the right order, each log write is accompanied by a \texttt{sfence} memory barrier. File systems that use redo/undo logging, such as NOVA or PMFS, use the tail pointer to revert changes in case of failure; hence, they update the tail pointer using expensive \texttt{clflush} and \texttt{sfence} instructions. In SplitFS, the tail pointer can be reconstructed in a crash, so there is no need to store it in PM. It is stored in DRAM and is atomically advanced using atomic Compare-and-Swap (\gls{cas}) operations, resulting in better parallel performance.
    \par
    The Strata file system uses a hybrid setup. In user space, operation logging is performed like SplitFS, except that Strata stores per-inode log pointers in PM. Eventually, the data contained in PM gets digested into block updates, which are then stored in redo logs located in the kernel digest areas. 
    \par
    \item The last trend is related to Shadow Paging. Although Shadow Paging avoids extensive in-place updates, it still suffers from high write amplification due to write propagation~\cite{condit-better-2009}. As discussed in \autoref{sec:file-system-design}, BPFS avoids write amplification by its \emph{Short-Circuit Shadow Paging}. However, it still incurs a large overhead when performing operations that cover a large part of the file system tree~\cite{xu-nova-2016}.\par
    Recent work aims to address this issue by introducing two new primitives that perform Shadow Paging without data movement. We already discussed the first primitive, \texttt{relink}, in~\autoref{sec:file-system-design}. Using \texttt{relink}, contiguous data regions can be moved atomically without any physical data movement, as seen in~\autoref{fig:usplit-relink}~\cite{kadekodi-splitfs-2019}. The second primitive, \texttt{pswap}, is implemented as a system call within ctFS's kernel space library, \emph{ctK}. Atomically, it swaps the Page Table Entries (\gls{pte}s) corresponding to two same-sized contiguous virtual memory regions in the Persistent Page Table (\gls{ppt}).
    \par
    We illustrate the purpose of \texttt{pswap} using an example. Suppose that a user wants to append to a file $x$ at offset $z$. First, \emph{ctU} (ctFS's user space library) allocates a \emph{staging} partition $P1$ in the PTT and copies the data to the same offset $z$ within this partition. Then, \emph{ctU} invokes \texttt{pswap} to atomically merge the original data contained in partition $P0$ with the new data in $P1$, persisting the data. Metadata updates are recorded in the redo log.
\end{inparaenum}

In short, we have seen four consistency techniques used in PM file systems: atomic in-place updates, log-structuring, shadow paging, and journaling. Most PM file systems implement a hybrid approach, as they identified that storing both data and metadata in a log results in poor performance. In the hybrid approach, data blocks are made consistent by performing inexpensive shadow paging using \texttt{relink} or \texttt{pswap} directives. Metadata persistence is achieved through a low-cost \emph{operation log}: a log that only stores file/directory operations, and not the actual data involved.


\subsection{Enforcing Write Ordering}

We already touched on the issue of \emph{write ordering} in the Introduction and Background (Sections~\ref{sec:introduction},~\ref{sec:background}). In short, the order in which data is written to disk may differ from the user's intentions, as a CPU can reorder writes for performance, resulting in data inconsistencies before entering the Persistence Domain (\gls{pd}). According to the relevant literature~\cite{condit-better-2009, bhandari-implications-2012, noauthor-intel-2022, luu-clwb-nodate}, there are four options to enforce write ordering. 

\begin{inparaenum}[(1)]
    \item The first option is to completely bypass the cache by performing \emph{write-through caching}: the cache and the actual PM location are written at the same time. In 2011, Bhandari et al.~\cite{bhandari-implications-2012} mentioned that this form of ordering has a slight advantage in CoW-based PM file systems. However, this claim should be taken with a grain of salt, as PM was not yet mainstream, thus platform support was minimal.
    \par
    \item Another option is to flush the entire cache at each memory barrier. A side effect is that the performance of other applications may degrade as its working set may be (partly) evicted from the cache.
    \par
    \item Alternatively, we can perform a more fine-grained flush. Instead of flushing the entire cache, we keep track of the cache lines in use and only flush those that contain file system (meta) data. Intel supports selective flushing by the \texttt{clflush} instruction. A \texttt{mfence} instruction ensures that all load and store instructions issued before \texttt{mfence} are serialized in the order they were performed. The \texttt{sfence} and \texttt{lfence} instructions provide this guarantee solely for store and load instructions, respectively~\cite{noauthor-intel-2022}.
    \par
    Although this form of ordering does not degrade the performance of the cache for other applications, it is still expensive. Bhandari et al.~\cite{bhandari-implications-2012} mention that one cache line flush takes around 300 CPU cycles on an Intel(R) Xeon(R) E5620 @ 2.4 GHz processor with a total of $12$ MB cache.\par
    To improve performance, Intel added two new instructions, namely \texttt{clflushopt} and \texttt{clwb}. The first provides an unordered version of \texttt{clflush}, allowing some concurrency when running multiple PM load/store instructions back-to-back. \texttt{clwb} behaves similarly to \texttt{clflushopt}, but does not invalidate the cache line~\cite{noauthor-persistent-2021, swift-hardware-2015, luu-clwb-nodate}.
    \par
    \item The final option is to allow software to explicitly communicate ordering constraints to hardware. The CPU is free to perform read and write caching, but must ensure that the ordering constraints are satisfied. In 2009, BPFS~\cite{condit-better-2009} proposed a new hardware extension called a \emph{epoch barrier}: a sequence of PM writes from the same thread delimited by a memory barrier issued in software. In 2015, Intel released this extension by introducing the \texttt{PCOMMIT} instruction. This instruction ensures that a user-specified memory range is written to persistent storage~\cite{luu-clwb-nodate}~\footnote{This instruction was later deprecated in favor of \texttt{clflushopt} and \texttt{clwb}~\cite{rudoff-deprecating-2016}}.
\end{inparaenum}

\autoref{tab:crash-consistency-techniques} displays an overview that relates the ordering techniques mentioned above to PM file systems. The file systems are ordered chronologically by their release date. One trend we can observe is that more recent work tends to use PM-specialized flush instructions, such as \texttt{clflushopt}, while older file systems use a more conventional \texttt{clflush} approach.

\par
In summary, we have seen multiple options to enforce write ordering: bypass the cache entirely, coarse/fine-grained flushing, and the \emph{epoah barrier}. Flushing the cache at cache line granularity using \texttt{clflushopt} is the preferred method of choice, as it allows for some concurrency when PM load/store instructions are executed.

\begin{table}[h]
    \centering
    \small
    \begin{tabular}{|l|P{0.32\linewidth}|P{0.32\linewidth}|}
      \hline
      & Metadata & Data Blocks \\
      \hline
      BPFS & \cellcolor{green!30}$\checkmark$ (4): \texttt{epochs}  & \cellcolor{green!30}$\checkmark$ (4): \texttt{epochs} \\
      \hline
      PMFS & \cellcolor{green!30}$\checkmark$ (3, 4) &  \cellcolor{red!30}$\xmark$ \\
      \hline
      HiNFS & \cellcolor{green!30}$\checkmark$ (3):  \texttt{clflush} \& \texttt{mfence} & \cellcolor{red!30}$\xmark$ \\
      \hline
      Ext4-DAX & \cellcolor{red!30}$\xmark$ & \cellcolor{red!30}$\xmark$ \\
      \hline
      SCMFS & \cellcolor{green!30}$\checkmark$ (3): \texttt{clflush} \& \texttt{mfence} & \cellcolor{green!30}$\checkmark$ (2) \\
      \hline
      SplitFS & \cellcolor{green!30}$\checkmark$ (3): \texttt{clflush} \& \texttt{sfence} & \cellcolor{green!30}$\checkmark$ (3): \texttt{clflush} \& \texttt{sfence} \\
      \hline
      Aerie & \cellcolor{green!30}$\checkmark$ (3): \texttt{clflush} \& \texttt{sfence} & \cellcolor{green!30}$\checkmark$ (3): \texttt{clflush} \& \texttt{sfence} \\
      \hline
      NOVA & \cellcolor{green!30}$\checkmark$ (3): \texttt{clflushopt} \& \texttt{clwb} & \cellcolor{green!30}$\checkmark$ (3): \texttt{clflushopt} \& \texttt{clwb} \\
      \hline
      Strata & \cellcolor{green!30}$\checkmark$ (3): \texttt{clflushopt}* or \texttt{clflush} & \cellcolor{green!30}$\checkmark$ (3): \texttt{clflushopt}* or \texttt{clflush} \\
      \hline
      Kuco & Not Specified & Not Specified \\
      \hline
      ZoFS &  \cellcolor{green!30}$\checkmark$ (3): \texttt{clflushopt} \& \texttt{clwb} & \cellcolor{green!30}$\checkmark$ (3): \texttt{clflushopt} \& \texttt{clwb} \\
      \hline
      HashFS & Not Applicable & Not Applicable \\
      \hline
      ByVFS & Not Applicable & Not Applicable \\
      \hline
    \end{tabular}
    \caption{PM file systems ordering enforcing techniques. *: preferred, if CPU support PMEM extensions}
    \label{tab:ordening-techniques}
\end{table}

\section{Open Problems and Future Work}

Although Persistent Memory brings exciting performance improvements to applications, widespread adoption has not (yet) been reached.
There seems to be no consensus among application developers about where to use PM and what benefits it provides~\cite{fellows-future-2021}. In addition, the best-performing file systems, Kuco, ZoFS and ctFS, require massive application refactoring as POSIX-like file semantics are not available. As a result, there is a trade-off to be made between user convenience and performance, which we have not covered in this work.
\par
In July 2022, Intel discontinued the Intel Optane product line. However, we do not expect software support to be discontinued in the near future; hence, PM-related research will continue. There are already emerging alternative devices: Kioxia and Everspin~\cite{sharwood-last-2022}. Another promising alternative is called the Compute Express Link (\gls{cxl}): a cache-coherent interconnect running on top of PCIe~\cite{xinyang-song-persistent-2022}. The early file system prototypes released by Microsoft~\cite{li-pond-2022} and Meta~\cite{maruf-tpp-2022} show promising results. Further work may include a study that looks at the advantages and differences of this technology compared to Intel's Optane PM. Other studies can dive into the specifics of implementing a CXL file system: “How CXL devices interact with modern workloads?", “What is the position of CXL memory inside the server micro-architecture?".

\section{Conclusion}

In this survey, we discussed how file systems address the three most prominent challenges when using Persistent Memory (\gls{pm}): the overhead shift from the device to the host I/O stack, indexing overhead, and data persistence.
\par
Before we answer the main research question, we first provide an answer for each of the subquestions:

\begin{itemize}
    \item \ref{qs:fs-design} - How have the properties/features of Persistent Memory led to changes in file system design?
    \par First, PM file systems bypass the page cache using Direct Access (DAX) to allow the user to directly modify file data in user space using MMU mappings. Other work focuses on the role of the kernel. In older PM file systems, the kernel bears full responsibility for maintaining metadata in a consistent state. More recent work shifts more responsibility to the user (see~\autoref{fig:pm-filesystems-timeline}), which in turn results in greater flexibility for applications to design custom-tailored file systems and extract the full potential of PM.
    
    \item \ref{qs:metadata-perf} - “Which optimizations help to decrease file indexing overhead in small I/O workloads?":
    \par The file indexing overhead is due to poor file mapping performance and the expensive VFS \emph{path walk}. The first issue is addressed by the introduction of a PM-optimized hash table. Another promising approach is to offload the translation to hardware via the Memory Management Unit.
    The performance impact of the VFS can be reduced by removing the redundant directory cache and introducing an adapted file path walker.
    \item \ref{qs:persistency} - “How can Persistent Memory file systems guarantee data crash consistency?":
    \par Data consistency must be enforced in software and hardware. In software, most file systems use a hybrid approach where data persistence is enforced through inexpensive Shadow Paging and metadata persistence through an \emph{operation log} and atomic in-place updates. At the hardware level, file systems use PM-optimized flushing instructions to enforce data ordering.
\end{itemize}

Now, we can answer the main research question, “Which file system design changes are needed to cope with the challenges that arise when using Persistent Memory?": We can definitely conclude that significant changes to the existing storage stack are necessary to extract the full performance potential of PM storage. When possible, the page cache should be avoided, as it serves no purpose in PM. Exposing device control to user space allows applications to fully leverage PM's byte-addressable properties. Data persistence must be enforced at the software and hardware level.



\bibliographystyle{plainurl}
\bibliography{main}

\begin{thebibliography}{10}

\bibitem{noauthor-rocksdb-nodate}
{RocksDB}: {A} {Persistent} {Key}-{Value} {Store} for {Flash} and {RAM}
  {Storage} {Optimized} (for {PMEM}).
\newblock URL: \url{https://github.com/pmem/pmem-rocksdb}.

\bibitem{noauthor-nvm-2017}
{NVM} {Programming} {Model} ({Version} 1.2), June 2017.
\newblock URL:
  \url{https://www.snia.org/sites/default/files/technical-work/npm/release/SNIA-NVM-Programming-Model-v1.2.pdf}.

\bibitem{noauthor-btrfs-2021}
btrfs - {A} modern copy on write({CoW}) filesystem for {Linux}, August 2021.
\newblock URL: \url{https://iceberg988.github.io/posts/btrfs/}.

\bibitem{noauthor-persistent-2021}
Persistent {Memory} {Extensions} - x86 - {WikiChip}, May 2021.
\newblock URL:
  \url{https://en.wikichip.org/wiki/x86/persistent\_memory\_extensions}.

\bibitem{noauthor-intel-2022}
Intel® 64 and {IA}-32 {Architectures} {Software} {Developer}’s {Manual},
  April 2022.
\newblock URL: \url{https://cdrdv2.intel.com/v1/dl/getContent/671200}.

\bibitem{agarwal-journaling-2020}
R.~Agarwal.
\newblock Journaling and {Log}-{Structured} {File} {Systems}, 2020.

\bibitem{bailey-operating-2011}
Katelin Bailey, Luis Ceze, Steven~D. Gribble, and Henry~M. Levy.
\newblock Operating system implications of fast, cheap, non-volatile memory.
\newblock In {\em Proceedings of the 13th {USENIX} conference on {Hot} topics
  in operating systems}, {HotOS}'13, page~2, USA, 2011. USENIX Association.

\bibitem{baldini-soares-flexsc-2010}
Livio Baldini~Soares and Michael Stumm.
\newblock “{FlexSC}: {Flexible} {System} {Call} {Scheduling} with
  {Exception}-{Less} {System} {Calls},”.
\newblock pages 33--46, January 2010.

\bibitem{bhandari-implications-2012}
K.~Bhandari, D.R. Chakrabarti, and H.-J Boehm.
\newblock Implications of {CPU} caching on byte-addressable non-volatile memory
  programming.
\newblock January 2012.

\bibitem{bhat-scaling-2017}
Srivatsa~S. Bhat, Rasha Eqbal, Austin~T. Clements, M.~Frans Kaashoek, and
  Nickolai Zeldovich.
\newblock Scaling a file system to many cores using an operation log.
\newblock In {\em Proceedings of the 26th {Symposium} on {Operating} {Systems}
  {Principles}}, {SOSP} '17, pages 69--86, New York, NY, USA, 2017. Association
  for Computing Machinery.
\newblock URL: \url{http://doi.org/10.1145/3132747.3132779}, \href
  {https://doi.org/10.1145/3132747.3132779}
  {\path{doi:10.1145/3132747.3132779}}.

\bibitem{braginsky-lock-free-2012}
Anastasia Braginsky and Erez Petrank.
\newblock A lock-free {B}+tree.
\newblock In {\em Proceedings of the twenty-fourth annual {ACM} symposium on
  {Parallelism} in algorithms and architectures}, {SPAA} '12, pages 58--67, New
  York, NY, USA, June 2012. Association for Computing Machinery.
\newblock URL: \url{http://doi.org/10.1145/2312005.2312016}, \href
  {https://doi.org/10.1145/2312005.2312016}
  {\path{doi:10.1145/2312005.2312016}}.

\bibitem{caheny-reducing-2018}
Paul Caheny, Lluc Alvarez, Said Derradji, Mateo Valero, Miquel Moretó, and
  Marc Casas.
\newblock Reducing {Cache} {Coherence} {Traffic} with a {NUMA}-{Aware}
  {Runtime} {Approach}.
\newblock {\em IEEE Transactions on Parallel and Distributed Systems},
  29(5):1174--1187, May 2018.
\newblock Conference Name: IEEE Transactions on Parallel and Distributed
  Systems.
\newblock \href {https://doi.org/10.1109/TPDS.2017.2787123}
  {\path{doi:10.1109/TPDS.2017.2787123}}.

\bibitem{cai-flatfs-2022}
Miao Cai.
\newblock {FlatFS}: {Flatten} {Hierarchical} {File} {System} {Namespace} on
  {Non}-volatile {Memories}.
\newblock {\em Usenix}, July 2022.
\newblock URL: \url{https://www.usenix.org/system/files/atc22-cai.pdf}.

\bibitem{cai-survey-2021}
Miao Cai and Hao Huang.
\newblock A survey of operating system support for persistent memory.
\newblock {\em Frontiers of Computer Science}, 15(4):154207, February 2021.
\newblock \href {https://doi.org/10.1007/s11704-020-9395-3}
  {\path{doi:10.1007/s11704-020-9395-3}}.

\bibitem{saarland-informatics-campus-cache-nodate}
Saarland~Informatics Campus.
\newblock Cache {Latencies}.
\newblock URL: \url{https://uops.info/cache.html}.

\bibitem{chen-understanding-2009}
Feng Chen, David~A. Koufaty, and Xiaodong Zhang.
\newblock Understanding intrinsic characteristics and system implications of
  flash memory based solid state drives.
\newblock {\em ACM SIGMETRICS Performance Evaluation Review}, 37(1):181--192,
  June 2009.
\newblock URL: \url{http://doi.org/10.1145/2492101.1555371}, \href
  {https://doi.org/10.1145/2492101.1555371}
  {\path{doi:10.1145/2492101.1555371}}.

\bibitem{chen-persistent-2015}
Shimin Chen and Qin Jin.
\newblock Persistent {B}+-trees in non-volatile main memory.
\newblock {\em Proceedings of the VLDB Endowment}, 8(7):786--797, February
  2015.
\newblock URL: \url{http://doi.org/10.14778/2752939.2752947}, \href
  {https://doi.org/10.14778/2752939.2752947}
  {\path{doi:10.14778/2752939.2752947}}.

\bibitem{chen-enabling-2018}
Shuo-Han Chen, Tseng-Yi Chen, Yuan-Hao Chang, Hsin-Wen Wei, and Wei-Kuan Shih.
\newblock Enabling union page cache to boost file access performance of
  {NVRAM}-based storage device.
\newblock In {\em Proceedings of the 55th {Annual} {Design} {Automation}
  {Conference}}, {DAC} '18, pages 1--6, New York, NY, USA, June 2018.
  Association for Computing Machinery.
\newblock URL: \url{http://doi.org/10.1145/3195970.3196045}, \href
  {https://doi.org/10.1145/3195970.3196045}
  {\path{doi:10.1145/3195970.3196045}}.

\bibitem{chen-kuco-2021}
Youmin Chen.
\newblock Kuco: {Scalable} {Persistent} {Memory} {File} {System} with
  {Kernel}-{Userspace} {Collaboration}.
\newblock {\em Proceedings of the 19th USENIX Conference on File and Storage
  Technologies.}, February 2021.
\newblock URL:
  \url{https://www.usenix.org/system/files/fast21-chen-youmin.pdf}.

\bibitem{comer-ubiquitous-1979}
Douglas Comer.
\newblock Ubiquitous {B}-{Tree}.
\newblock {\em ACM Computing Surveys}, 11(2):121--137, June 1979.
\newblock URL: \url{http://doi.org/10.1145/356770.356776}, \href
  {https://doi.org/10.1145/356770.356776} {\path{doi:10.1145/356770.356776}}.

\bibitem{kernel-development-community-pathname-nodate}
Kernel~Development Community.
\newblock Pathname lookup — {The} {Linux} {Kernel} documentation.
\newblock Kernel version 6.1.0.
\newblock URL:
  \url{https://www.kernel.org/doc/html/latest/filesystems/path-lookup.html}.

\bibitem{condit-better-2009}
Jeremy Condit, Edmund~B. Nightingale, Christopher Frost, Engin Ipek, Benjamin
  Lee, Doug Burger, and Derrick Coetzee.
\newblock Better {I}/{O} through byte-addressable, persistent memory.
\newblock In {\em Proceedings of the {ACM} {SIGOPS} 22nd symposium on
  {Operating} systems principles - {SOSP} '09}, page 133, Big Sky, Montana,
  USA, 2009. ACM Press.
\newblock URL: \url{http://portal.acm.org/citation.cfm?doid=1629575.1629589},
  \href {https://doi.org/10.1145/1629575.1629589}
  {\path{doi:10.1145/1629575.1629589}}.

\bibitem{copeland-modern-2020}
B.~Jack Copeland.
\newblock The {Modern} {History} of {Computing}.
\newblock In Edward~N. Zalta, editor, {\em The {Stanford} {Encyclopedia} of
  {Philosophy}}. Metaphysics Research Lab, Stanford University, winter 2020
  edition, 2020.
\newblock URL:
  \url{https://plato.stanford.edu/archives/win2020/entries/computing-history/}.

\bibitem{corbet-memory-2015}
Jonathan Corbet.
\newblock Memory {Protection} {Keys}, May 2015.
\newblock URL: \url{https://lwn.net/Articles/643797/}.

\bibitem{cox-jedec-nodate}
Alvin Cox.
\newblock {JEDEC} {SSD} {Specifications} {Explained}.
\newblock URL:
  \url{https://www.jedec.org/sites/default/files/Alvin\_Cox%20[Compatibility%20Mode]\_0.pdf}.

\bibitem{debnath-flashstore-2010}
Biplob Debnath, Sudipta Sengupta, and Jin Li.
\newblock {FlashStore}: high throughput persistent key-value store.
\newblock {\em Proceedings of the VLDB Endowment}, 3(1-2):1414--1425, September
  2010.
\newblock URL: \url{https://dl.acm.org/doi/10.14778/1920841.1921015}, \href
  {https://doi.org/10.14778/1920841.1921015}
  {\path{doi:10.14778/1920841.1921015}}.

\bibitem{dong-performance-2019}
Mingkai Dong, Heng Bu, Jifei Yi, Benchao Dong, and Haibo Chen.
\newblock Performance and protection in the {ZoFS} user-space {NVM} file
  system.
\newblock In {\em Proceedings of the 27th {ACM} {Symposium} on {Operating}
  {Systems} {Principles}}, {SOSP} '19, pages 478--493, New York, NY, USA, 2019.
  Association for Computing Machinery.
\newblock URL: \url{http://doi.org/10.1145/3341301.3359637}, \href
  {https://doi.org/10.1145/3341301.3359637}
  {\path{doi:10.1145/3341301.3359637}}.

\bibitem{dong-rocksdb-2021}
Siying Dong, Andrew Kryczka, Yanqin Jin, and Michael Stumm.
\newblock {RocksDB}: {Evolution} of {Development} {Priorities} in a {Key}-value
  {Store} {Serving} {Large}-scale {Applications}.
\newblock {\em ACM Transactions on Storage}, 17(4):26:1--26:32, October 2021.
\newblock URL: \url{http://doi.org/10.1145/3483840}, \href
  {https://doi.org/10.1145/3483840} {\path{doi:10.1145/3483840}}.

\bibitem{dulloor-pmfs-2014}
Subramanya~R. Dulloor, Sanjay Kumar, Anil Keshavamurthy, Philip Lantz, Dheeraj
  Reddy, Rajesh Sankaran, and Jeff Jackson.
\newblock {PMFS}: {System} software for persistent memory.
\newblock In {\em Proceedings of the {Ninth} {European} {Conference} on
  {Computer} {Systems}}, {EuroSys} '14, pages 1--15, New York, NY, USA, April
  2014. Association for Computing Machinery.
\newblock URL: \url{http://doi.org/10.1145/2592798.2592814}, \href
  {https://doi.org/10.1145/2592798.2592814}
  {\path{doi:10.1145/2592798.2592814}}.

\bibitem{fellows-future-2021}
Russell Fellows.
\newblock The {Future} of {Optane} and {Persistent} {Memory}, March 2021.
\newblock URL:
  \url{https://www.linkedin.com/pulse/future-optane-persistent-memory-russell-fellows-1e}.

\bibitem{fingerhut-does-2014}
Steve Fingerhut.
\newblock Does {Storage} break {Moore}’s {Law}? {A} {Look} at {SSD} vs {HDD},
  July 2014.
\newblock URL:
  \url{https://blog.westerndigital.com/does-storage-break-moores-law/}.

\bibitem{fontana-moores-2018}
Robert~E. Fontana and Gary~M. Decad.
\newblock Moore’s law realities for recording systems and memory storage
  components.
\newblock {\em AIP Advances}, 8(5):056506, May 2018.
\newblock Publisher: American Institute of Physics.
\newblock URL: \url{http://aip.scitation.org/doi/10.1063%2F1.5007621}, \href
  {https://doi.org/10.1063/1.5007621} {\path{doi:10.1063/1.5007621}}.

\bibitem{linux-foundation-direct-nodate}
Linux Foundation.
\newblock Direct {Access} for files.
\newblock URL:
  \url{https://www.kernel.org/doc/Documentation/filesystems/dax.txt}.

\bibitem{linux-foundation-ext4-nodate}
Linux Foundation.
\newblock ext4 {Data} {Structures} and {Algorithms} — {The} {Linux} {Kernel}
  documentation.
\newblock URL:
  \url{https://www.kernel.org/doc/html/latest/filesystems/ext4/globals.html#super-block}.

\bibitem{linux-foundation-page-nodate}
Linux Foundation.
\newblock Page {Table} {Management}.
\newblock URL:
  \url{https://www.kernel.org/doc/gorman/html/understand/understand006.html}.

\bibitem{gu-pisces-2019}
Jinyu Gu, Qianqian Yu, Xiayang Wang, Zhaoguo Wang, Binyu Zang, Haibing Guan,
  and Haibo Chen.
\newblock Pisces: a scalable and efficient persistent transactional memory.
\newblock In {\em Proceedings of the 2019 {USENIX} {Conference} on {Usenix}
  {Annual} {Technical} {Conference}}, {USENIX} {ATC} '19, pages 913--928, USA,
  July 2019. USENIX Association.

\bibitem{gugnani-understanding-2021}
Shashank Gugnani, Arjun Kashyap, and Xiaoyi Lu.
\newblock Understanding the idiosyncrasies of real persistent memory.
\newblock {\em Proceedings of the VLDB Endowment}, 14(4):626--639, February
  2021.
\newblock URL: \url{http://doi.org/10.14778/3436905.3436921}, \href
  {https://doi.org/10.14778/3436905.3436921}
  {\path{doi:10.14778/3436905.3436921}}.

\bibitem{iaik-paging-nodate}
IAIK.
\newblock Paging on {Intel} x86-64 – {IAIK}.
\newblock URL:
  \url{https://www.iaik.tugraz.at/teaching/materials/os/tutorials/paging-on-intel-x86-64/}.

\bibitem{izraelevitz-basic-2019}
Joseph Izraelevitz, Jian Yang, Lu~Zhang, Juno Kim, Xiao Liu, Amirsaman
  Memaripour, Yun~Joon Soh, Zixuan Wang, Yi~Xu, Subramanya~R. Dulloor, Jishen
  Zhao, and Steven Swanson.
\newblock Basic {Performance} {Measurements} of the {Intel} {Optane} {DC}
  {Persistent} {Memory} {Module}, August 2019.
\newblock arXiv:1903.05714 [cs].
\newblock URL: \url{http://arxiv.org/abs/1903.05714}, \href
  {https://doi.org/10.48550/arXiv.1903.05714}
  {\path{doi:10.48550/arXiv.1903.05714}}.

\bibitem{kadekodi-splitfs-2019}
Rohan Kadekodi, Se~Kwon Lee, Sanidhya Kashyap, Taesoo Kim, Aasheesh Kolli, and
  Vijay Chidambaram.
\newblock {SplitFS}: reducing software overhead in file systems for persistent
  memory.
\newblock In {\em Proceedings of the 27th {ACM} {Symposium} on {Operating}
  {Systems} {Principles}}, pages 494--508, Huntsville Ontario Canada, October
  2019. ACM.
\newblock URL: \url{https://dl.acm.org/doi/10.1145/3341301.3359631}, \href
  {https://doi.org/10.1145/3341301.3359631}
  {\path{doi:10.1145/3341301.3359631}}.

\bibitem{koo-modernizing-2021}
Jinhyung Koo, Junsu Im, Jooyoung Song, Juhyung Park, Eunji Lee, Bryan~S. Kim,
  and Sungjin Lee.
\newblock Modernizing {File} {System} through {In}-{Storage} {Indexing}.
\newblock {ODSI} '21, pages 75--92, 2021.
\newblock URL: \url{https://www.usenix.org/conference/osdi21/presentation/koo}.

\bibitem{kwon-strata-2017}
Youngjin Kwon, Henrique Fingler, Tyler Hunt, Simon Peter, Emmett Witchel, and
  Thomas Anderson.
\newblock Strata: {A} {Cross} {Media} {File} {System}.
\newblock In {\em Proceedings of the 26th {Symposium} on {Operating} {Systems}
  {Principles}}, pages 460--477, Shanghai China, October 2017. ACM.
\newblock URL: \url{https://dl.acm.org/doi/10.1145/3132747.3132770}, \href
  {https://doi.org/10.1145/3132747.3132770}
  {\path{doi:10.1145/3132747.3132770}}.

\bibitem{lee-f2fs-2015}
Changman Lee, Dongho Sim, Jooyoung Hwang, and Sangyeun Cho.
\newblock \{{F2FS}\}: {A} {New} {File} {System} for {Flash} {Storage}.
\newblock 13th {USENIX} {Conference} on {File} and {Storage} {Technologies}
  ({FAST} 15), pages 273--286, 2015.
\newblock URL:
  \url{https://www.usenix.org/conference/fast15/technical-sessions/presentation/lee}.

\bibitem{leventhal-flash-2008}
Adam Leventhal.
\newblock Flash storage memory.
\newblock {\em Communications of the ACM}, 51(7):47--51, July 2008.
\newblock URL: \url{http://doi.org/10.1145/1364782.1364796}, \href
  {https://doi.org/10.1145/1364782.1364796}
  {\path{doi:10.1145/1364782.1364796}}.

\bibitem{li-pond-2022}
Huaicheng Li, Daniel~S. Berger, Stanko Novakovic, Lisa Hsu, Dan Ernst, Pantea
  Zardoshti, Monish Shah, Samir Rajadnya, Scott Lee, Ishwar Agarwal, Mark~D.
  Hill, Marcus Fontoura, and Ricardo Bianchini.
\newblock Pond: {CXL}-{Based} {Memory} {Pooling} {Systems} for {Cloud}
  {Platforms}.
\newblock {\em ACM International Conference on Architectural Support for
  Programming Languages and Operating Systems (ASPLOS ’23) (}, October 2022.
\newblock arXiv:2203.00241 [cs].
\newblock URL: \url{http://arxiv.org/abs/2203.00241}.

\bibitem{li-ctfs-2022}
Ruibin Li.
\newblock {ctFS}: {Replacing} {File} {Indexing} with {Hardware} {Memory}
  {Translation} through {Contiguous} {File} {Allocation} for {Persistent}
  {Memory}.
\newblock {\em 20th USENIX Conference on File and Storage Technologies (FAST
  22)}, 2022.
\newblock URL: \url{https://www.usenix.org/conference/fast22/presentation/li}.

\bibitem{lu-loose-ordering-2014}
Youyou Lu, Jiwu Shu, Long Sun, and Onur Mutlu.
\newblock Loose-{Ordering} {Consistency} for persistent memory.
\newblock In {\em 2014 {IEEE} 32nd {International} {Conference} on {Computer}
  {Design} ({ICCD})}, pages 216--223, October 2014.
\newblock ISSN: 1063-6404.
\newblock \href {https://doi.org/10.1109/ICCD.2014.6974684}
  {\path{doi:10.1109/ICCD.2014.6974684}}.

\bibitem{luu-clwb-nodate}
Dan Luu.
\newblock {CLWB} and {PCOMMIT}.
\newblock URL: \url{https://danluu.com/clwb-pcommit/}.

\bibitem{marco-gisbert-address-2019}
Hector Marco-Gisbert and Ismael Ripoll~Ripoll.
\newblock Address {Space} {Layout} {Randomization} {Next} {Generation}.
\newblock {\em Applied Sciences}, 9(14):2928, January 2019.
\newblock Number: 14 Publisher: Multidisciplinary Digital Publishing Institute.
\newblock URL: \url{https://www.mdpi.com/2076-3417/9/14/2928}, \href
  {https://doi.org/10.3390/app9142928} {\path{doi:10.3390/app9142928}}.

\bibitem{maruf-tpp-2022}
Hasan~Al Maruf, Hao Wang, Abhishek Dhanotia, Johannes Weiner, Niket Agarwal,
  Pallab Bhattacharya, Chris Petersen, Mosharaf Chowdhury, Shobhit Kanaujia,
  and Prakash Chauhan.
\newblock {TPP}: {Transparent} {Page} {Placement} for {CXL}-{Enabled} {Tiered}
  {Memory}, June 2022.
\newblock arXiv:2206.02878 [cs].
\newblock URL: \url{http://arxiv.org/abs/2206.02878}.

\bibitem{mathur-new-2007}
Avantika Mathur, Mingming Cao, Suparna Bhattacharya, Andreas Dilger, Alex
  Tomas, and Laurent Vivier.
\newblock The new ext4 filesystem: {Current} status and future plans.
\newblock {\em Proceedings of the Linux Symposium}, January 2007.

\bibitem{mcdougall-filebench-2004}
Richard McDougall.
\newblock {FileBench}, 2004.
\newblock URL:
  \url{http://www.nfsv4bat.org/Documents/nasconf/2004/filebench.pdf}.

\bibitem{mckenney-memory-2005}
Paul McKenney.
\newblock Memory {Ordering} in {Modern} {Microprocessors}, {Part} {I}
  {\textbar} {Linux} {Journal}.
\newblock {\em Linux Journal}, June 2005.
\newblock URL: \url{https://www.linuxjournal.com/article/8211}.

\bibitem{min-sfs-2012}
Changwoo Min.
\newblock {SFS}: {Random} {Write} {Considered} {Harmful} in {Solid} {State}
  {Drives}.
\newblock In {\em 10th {USENIX} {Conference} on {File} and {Storage}
  {Technologies} ({FAST} 12)}, 2012.
\newblock URL:
  \url{https://www.usenix.org/conference/fast12/sfs-random-write-considered-harmful-solid-state-drives}.

\bibitem{mohan-analyzing-2017}
Jayashree Mohan, Rohan Kadekodi, and Vijay Chidambaram.
\newblock Analyzing {IO} {Amplification} in {Linux} {File} {Systems}, July
  2017.
\newblock arXiv:1707.08514 [cs].
\newblock URL: \url{http://arxiv.org/abs/1707.08514}, \href
  {https://doi.org/10.48550/arXiv.1707.08514}
  {\path{doi:10.48550/arXiv.1707.08514}}.

\bibitem{morris-data-nodate}
John Morris.
\newblock Data {Structures} and {Algorithms}: {Red}-{Black} {Trees}.
\newblock URL:
  \url{https://www.cs.auckland.ac.nz/software/AlgAnim/red\_black.html}.

\bibitem{neal-hashfs-2021}
Ian Neal.
\newblock {HashFS}: {Rethinking} {File} {Mapping} for {Persistent} {Memory}.
\newblock {\em Proceedings of the 19th USENIX Conference on File and Storage
  Technologies}, February 2021.
\newblock URL: \url{https://www.usenix.org/system/files/fast21-neal.pdf}.

\bibitem{ou-hinfs-2016}
Jiaxin Ou, Jiwu Shu, and Youyou Lu.
\newblock {HiNFS}: {A} high performance file system for non-volatile main
  memory.
\newblock In {\em Proceedings of the {Eleventh} {European} {Conference} on
  {Computer} {Systems}}, {EuroSys} '16, pages 1--16, New York, NY, USA, April
  2016. Association for Computing Machinery.
\newblock URL: \url{http://doi.org/10.1145/2901318.2901324}, \href
  {https://doi.org/10.1145/2901318.2901324}
  {\path{doi:10.1145/2901318.2901324}}.

\bibitem{oneil-log-structured-1996}
Patrick O’Neil, Edward Cheng, Dieter Gawlick, and Elizabeth O’Neil.
\newblock The log-structured merge-tree ({LSM}-tree).
\newblock {\em Acta Informatica}, 33(4):351--385, June 1996.
\newblock \href {https://doi.org/10.1007/s002360050048}
  {\path{doi:10.1007/s002360050048}}.

\bibitem{pagh-cuckoo-2004}
Rasmus Pagh and Flemming~Friche Rodler.
\newblock Cuckoo hashing.
\newblock {\em Journal of Algorithms}, 51(2):122--144, May 2004.
\newblock URL:
  \url{https://www.sciencedirect.com/science/article/pii/S0196677403001925},
  \href {https://doi.org/10.1016/j.jalgor.2003.12.002}
  {\path{doi:10.1016/j.jalgor.2003.12.002}}.

\bibitem{peng-system-2019}
Ivy~B. Peng, Maya~B. Gokhale, and Eric~W. Green.
\newblock System evaluation of the {Intel} optane byte-addressable {NVM}.
\newblock In {\em Proceedings of the {International} {Symposium} on {Memory}
  {Systems}}, {MEMSYS} '19, pages 304--315, New York, NY, USA, September 2019.
  Association for Computing Machinery.
\newblock URL: \url{http://doi.org/10.1145/3357526.3357568}, \href
  {https://doi.org/10.1145/3357526.3357568}
  {\path{doi:10.1145/3357526.3357568}}.

\bibitem{rudoff-deprecating-2016}
Andy Rudoff.
\newblock Deprecating the {PCOMMIT} {Instruction}, September 2016.
\newblock URL:
  \url{https://www.intel.com/content/www/us/en/developer/articles/technical/deprecate-pcommit-instruction.html}.

\bibitem{sauthoff-costs-2021}
Georg Sauthoff.
\newblock On the {Costs} of {Syscalls}, August 2021.
\newblock URL: \url{https://gms.tf/on-the-costs-of-syscalls.html}.

\bibitem{seltzer-implementation-1993}
Margo Seltzer, Keith Bostic, Marshall McKusick, and Carl Staelin.
\newblock An {Implementation} of a {Log}-{Structured} {File} {System} for
  {UNIX}.
\newblock {USENIX} '1993, pages 307--326, January 1993.

\bibitem{sharwood-last-2022}
Simon Sharwood.
\newblock Last week {Intel} killed {Optane}. {Competing} tech keeps coming.
\newblock {\em The A Register}, February 2022.
\newblock URL:
  \url{https://www.theregister.com/2022/08/02/kioxia\_everspin\_persistent\_memory/}.

\bibitem{spooner-intel-2006}
John~G. Spooner.
\newblock Intel {Previews} {Potential} {Replacement} for {Flash} {Memory}.
\newblock {\em eWEEK}, September 2006.
\newblock URL:
  \url{https://www.eweek.com/pc-hardware/intel-previews-potential-replacement-for-flash-memory/}.

\bibitem{su-survey-2021}
Chao Su and Qingkai Zeng.
\newblock Survey of {CPU} {Cache}-{Based} {Side}-{Channel} {Attacks}:
  {Systematic} {Analysis}, {Security} {Models}, and {Countermeasures}.
\newblock {\em Security and Communication Networks}, 2021:e5559552, June 2021.
\newblock Publisher: Hindawi.
\newblock URL: \url{https://www.hindawi.com/journals/scn/2021/5559552/}, \href
  {https://doi.org/10.1155/2021/5559552} {\path{doi:10.1155/2021/5559552}}.

\bibitem{swanson-early-2019}
Steven Swanson.
\newblock Early {Measurements} of {Intel}’s {3DXPoint} {Persistent} {Memory}
  {DIMMs}, April 2019.
\newblock URL:
  \url{https://www.sigarch.org/early-measurements-of-intels-3dxpoint-persistent-memory-dimms/}.

\bibitem{swift-hardware-2015}
Micheal Swift.
\newblock Hardware {Support} for {NVM} {Programming}, March 2015.
\newblock URL:
  \url{https://research.cs.wisc.edu/sonar/tutorial/03-hardware.pdf}.

\bibitem{tianhua-design-2008}
Liu Tianhua, Zhu Hongfeng, Chang Guiran, and Zhou Chuansheng.
\newblock The {Design} and {Implementation} of {Zero}-{Copy} for {Linux}.
\newblock In {\em 2008 {Eighth} {International} {Conference} on {Intelligent}
  {Systems} {Design} and {Applications}}, volume~1, pages 121--126, November
  2008.
\newblock ISSN: 2164-7151.
\newblock \href {https://doi.org/10.1109/ISDA.2008.102}
  {\path{doi:10.1109/ISDA.2008.102}}.

\bibitem{volos-aerie-2014}
Haris Volos, Sanketh Nalli, Sankarlingam Panneerselvam, Venkatanathan
  Varadarajan, Prashant Saxena, and Michael~M. Swift.
\newblock Aerie: flexible file-system interfaces to storage-class memory.
\newblock In {\em Proceedings of the {Ninth} {European} {Conference} on
  {Computer} {Systems}}, {EuroSys} '14, pages 1--14, New York, NY, USA, April
  2014. Association for Computing Machinery.
\newblock URL: \url{http://doi.org/10.1145/2592798.2592810}, \href
  {https://doi.org/10.1145/2592798.2592810}
  {\path{doi:10.1145/2592798.2592810}}.

\bibitem{wan-empirical-2016}
Hu~Wan, Youyou Lu, Yuanchao Xu, and Jiwu Shu.
\newblock Empirical study of redo and undo logging in persistent memory.
\newblock In {\em 2016 5th {Non}-{Volatile} {Memory} {Systems} and
  {Applications} {Symposium} ({NVMSA})}, pages 1--6, August 2016.
\newblock \href {https://doi.org/10.1109/NVMSA.2016.7547178}
  {\path{doi:10.1109/NVMSA.2016.7547178}}.

\bibitem{wang-byvfs-2018}
Ying Wang.
\newblock {ByVFS}: {Caching} or {Not}: {Rethinking} {Virtual} {File} {System}
  for {Non}-{Volatile} {Main} {Memory}.
\newblock September 2018.
\newblock URL:
  \url{https://www.usenix.org/system/files/conference/hotstorage18/hotstorage18-paper-wang.pdf}.

\bibitem{waymann-survey-2017}
Lukas Waymann.
\newblock A {Survey} of {CPU} {Caches}, October 2017.
\newblock URL: \url{https://meribold.org/2017/10/20/survey-of-cpu-caches/}.

\bibitem{wohlin-guidelines-2014}
Claes Wohlin.
\newblock Guidelines for snowballing in systematic literature studies and a
  replication in software engineering.
\newblock In {\em Proceedings of the 18th {International} {Conference} on
  {Evaluation} and {Assessment} in {Software} {Engineering}}, {EASE} '14, pages
  1--10, New York, NY, USA, 2014. Association for Computing Machinery.
\newblock URL: \url{http://doi.org/10.1145/2601248.2601268}, \href
  {https://doi.org/10.1145/2601248.2601268}
  {\path{doi:10.1145/2601248.2601268}}.

\bibitem{wright-extending-2007}
Charles~P. Wright, Richard Spillane, Gopalan Sivathanu, and Erez Zadok.
\newblock Extending {ACID} semantics to the file system.
\newblock {\em ACM Transactions on Storage}, 3(2):4--es, June 2007.
\newblock URL: \url{http://doi.org/10.1145/1242520.1242521}, \href
  {https://doi.org/10.1145/1242520.1242521}
  {\path{doi:10.1145/1242520.1242521}}.

\bibitem{wu-scmfs-2013}
Xiaojian Wu, Sheng Qiu, and A.~L. Narasimha~Reddy.
\newblock {SCMFS}: {A} {File} {System} for {Storage} {Class} {Memory} and its
  {Extensions}.
\newblock {\em ACM Transactions on Storage}, 9(3):7:1--7:23, August 2013.
\newblock URL: \url{http://doi.org/10.1145/2501620.2501621}, \href
  {https://doi.org/10.1145/2501620.2501621}
  {\path{doi:10.1145/2501620.2501621}}.

\bibitem{xinyang-song-persistent-2022}
{Xinyang, Song} and Sihang Liu.
\newblock Persistent {Memory} – {A} {New} {Hope}, September 2022.
\newblock URL: \url{https://www.sigarch.org/persistent-memory-a-new-hope/}.

\bibitem{xu-nova-2016}
Jian Xu and Steven Swanson.
\newblock {NOVA}: {A} {Log}-structured {File} {System} for {Hybrid}
  \{{Volatile}/{Non}-volatile\} {Main} {Memories}.
\newblock {FAST} '16, pages 323--338, 2016.
\newblock URL:
  \url{https://www.usenix.org/conference/fast16/technical-sessions/presentation/xu}.

\bibitem{yang-empirical-nodate}
Jian Yang.
\newblock An {Empirical} {Guide} to the {Behavior} and {Use} of {Scalable}
  {Persistent} {Memory}.
\newblock {\em Proceedings of the 18th USENIX Conference on File and Storage
  Technologies (FAST ’20)}.
\newblock URL: \url{https://www.usenix.org/system/files/fast20-yang.pdf}.

\bibitem{yang-i-cash-2011}
Qing Yang and Jin Ren.
\newblock I-{CASH}: {Intelligently} {Coupled} {Array} of {SSD} and {HDD}.
\newblock In {\em 2011 {IEEE} 17th {International} {Symposium} on {High}
  {Performance} {Computer} {Architecture}}, pages 278--289, February 2011.
\newblock ISSN: 2378-203X.
\newblock \href {https://doi.org/10.1109/HPCA.2011.5749736}
  {\path{doi:10.1109/HPCA.2011.5749736}}.

\bibitem{yang-spmfs-2021}
Yang Yang, Qiang Cao, Jie Yao, Yuanyuan Dong, and Weikang Kong.
\newblock {SPMFS}: {A} {Scalable} {Persistent} {Memory} {File} {System} on
  {Optane} {Persistent} {Memory}.
\newblock In {\em 50th {International} {Conference} on {Parallel}
  {Processing}}, {ICPP} 2021, pages 1--10, New York, NY, USA, October 2021.
  Association for Computing Machinery.
\newblock URL: \url{http://doi.org/10.1145/3472456.3472503}, \href
  {https://doi.org/10.1145/3472456.3472503}
  {\path{doi:10.1145/3472456.3472503}}.

\bibitem{zhang-study-2015}
Yiying Zhang and Steven Swanson.
\newblock A study of application performance with non-volatile main memory.
\newblock In {\em 2015 31st {Symposium} on {Mass} {Storage} {Systems} and
  {Technologies} ({MSST})}, pages 1--10, May 2015.
\newblock ISSN: 2160-1968.
\newblock \href {https://doi.org/10.1109/MSST.2015.7208275}
  {\path{doi:10.1109/MSST.2015.7208275}}.

\end{thebibliography}


\end{document}